\documentclass{emulateapj}

\shorttitle{Modeling of the dynamics of $z\sim2$ galaxies}
\shortauthors{Cresci et al.}

\begin{document}


\title{The SINS survey: modeling the dynamics of $z\sim2$ galaxies and the 
   high-z Tully-Fisher relation}\altaffiltext{$\star$}{Based on observations 
   obtained at the Very Large Telescope (VLT) 
   of the European Southern Observatory, Paranal, Chile in the context of 
   guaranteed time programs 073.B-9018, 074.A-9011, 075.A-0466, 076.A-0527, 
   077.A-0576, 078.A-0600, 078.A-0055, 079.A-0341, 080.A-0330 and 080.A-0635.}


\author{G. Cresci\altaffilmark{1}, E.~K.~S. Hicks\altaffilmark{1}, R. Genzel\altaffilmark{1,2}, 
N.~M. F\"orster Schreiber\altaffilmark{1}, R. Davies\altaffilmark{1}, N. Bouch\'e\altaffilmark{1}, 
P. Buschkamp\altaffilmark{1}, S. Genel\altaffilmark{1}, K. Shapiro\altaffilmark{3},  
L. Tacconi\altaffilmark{1}, J. Sommer-Larsen\altaffilmark{4,5}, A. Burkert\altaffilmark{6}, 
F. Eisenhauer\altaffilmark{1}, O. Gerhard\altaffilmark{1}, D. Lutz\altaffilmark{1}, T. Naab\altaffilmark{6}, 
A. Sternberg\altaffilmark{7}, A. Cimatti\altaffilmark{8}, E. Daddi\altaffilmark{9}, D.~K. Erb\altaffilmark{10}, 
J. Kurk\altaffilmark{11}, S.~L. Lilly\altaffilmark{12}, A. Renzini\altaffilmark{13}, A. Shapley\altaffilmark{14}, 
C.~C. Steidel\altaffilmark{15} and K. Caputi\altaffilmark{12}
}
%\affil{Max-Planck-Institut f\"ur extraterrestrische Physik,
%        Postfach 1312, D-85741 Garching, Germany}
%\email{gcresci@mpe.mpg.de}
%\author{K. Shapiro}
%\affil{Department of Astronomy, Campbell Hall, University of California, Berkeley, CA 94720, USA}
%\author{J. Sommer-Larsen\altaffilmark{2}}
%\affil{Excellence Cluster Universe, Technische Universit\"at M\"unchen; Boltzmanstr. 2, D-85748 Garching, Germany}
%\author{A. Cimatti}
%\affil{Dipartimento di Astronomia, Universit\'a degli Studi di Bologna, via Ranzani 1, I-40127 Bologna, Italy}
%\author{E. Daddi}2007ApJ...660...97C
%\affil{CEA, Laboratoire AIM - CNRS - Universit\'e Paris Diderot, Irfu/SAp, Orme des Merisiers, F-91191 Gif-sur-Yvette, France}
%\author{S.~L. Lilly}
%\affil{Institute of Astronomy, Department of Physics, Eidgen\"ossische Technische Hochshule, ETH Z\"urich, CH-8093, Switzerland}
%\and
%\author{A. Renzini}
%\affil{INAF - Osservatorio Astronomico di Padova, vicolo dell'Osservatorio 5, I-35122 Padova, Italy}


\altaffiltext{1}{\small Max-Planck-Institut f\"ur extraterrestrische Physik, Postfach 1312, D-85741 Garching, Germany (gcresci@mpe.mpg.de)}
\altaffiltext{2}{\small Department of Physics, Campbell Hall, University of California, Berkeley, CA 94720, USA}
\altaffiltext{3}{\small Department of Astronomy, Campbell Hall, University of California, Berkeley, CA 94720, USA}
\altaffiltext{4}{\small Excellence Cluster Universe, Technische Universit\"at M\"unchen; Boltzmanstr. 2, D-85748 Garching, Germany}
\altaffiltext{5}{\small Dark Cosmology Centre, Niels Bohr Institute, University of Copenhagen, Juliane Marie Vej 30, 2100 Copenhagen, Denmark}
\altaffiltext{6}{\small Universit\"ats-Sternwarte Ludwig-Maximilians Universit\"at (USM), Scheinerstr. 1, M\"unchen, D-81679, Germany}
\altaffiltext{7}{\small School of Physics and Astronomy, Tel Aviv University, Tel Aviv 69978, Israel}
\altaffiltext{8}{\small Dipartimento di Astronomia, Universit\'a degli Studi di Bologna, via Ranzani 1, I-40127 Bologna, Italy}
\altaffiltext{9}{\small CEA, Laboratoire AIM - CNRS - Universit\'e Paris Diderot, Irfu/SAp, Orme des Merisiers, F-91191 Gif-sur-Yvette, France}
\altaffiltext{10}{\small Harvard-Smithsonian Center for Astrophysics, 60 Garden Street, Cambridge, Mass. 02138, USA}
\altaffiltext{11}{\small Max-Planck-Institut f\"ur Astronomie, K\"onigstuhl 17, 69117 Heidelberg, Germany}
\altaffiltext{12}{\small Institute of Astronomy, Department of Physics, Eidgen\"ossische Technische Hochshule, ETH Z\"urich, CH-8093, Switzerland}
\altaffiltext{13}{\small INAF - Osservatorio Astronomico di Padova, vicolo dell'Osservatorio 5, I-35122 Padova, Italy}
\altaffiltext{14}{\small Department of Physics and Astronomy, 430 Portola Plaza, University of California, Los Angeles, CA 90095-1547, USA}
\altaffiltext{15}{\small California Institute of Technology, MS 105-24, Pasadena, CA 91125, USA}

\begin{abstract}
We present the modeling of SINFONI integral field dynamics of 18 star forming galaxies 
at $z\sim2$  from H$\alpha$ line emission. The galaxies are selected from 
the larger sample of the SINS survey, based on the prominence of 
ordered rotational motions with respect to more complex merger induced dynamics.
The quality of the data allows us to carefully select systems with kinematics dominated by 
rotation, and to model the gas dynamics across the whole galaxy using suitable  
exponential disk models. 
We obtain a good correlation between the dynamical mass and the stellar mass, 
finding that large gas fractions ($M_{gas}\approx M_*$) are required to 
explain the difference between the two quantities. 
We use the derived stellar mass and maximum rotational velocity $V_{max}$ from the modeling 
to construct for the first time the stellar mass Tully-Fisher relation at $z\sim2.2$. 
The relation obtained shows a slope similar to what is observed at lower redshift, but we detect 
an evolution of the zero point. We find that at $z\sim2.2$ there is an offset in $log(M_*)$ 
for a given rotational velocity of $0.41\pm0.11$ with respect to the local Universe.
This result is consistent with the predictions of the latest N-body/hydrodynamical 
simulations of disk formation and evolution, which invoke gas accretion onto the forming disk 
in filaments and cooling flows. This scenario is in agreement with other dynamical evidence from 
SINS, where gas accretion from the halo is required to reproduce the observed properties 
of a large fraction  of the $z\sim2$ galaxies. 
\end{abstract}





\section{Introduction}

%The formation of massive galaxies is a key issue of current theoretical 
%and observational research that can now be tackled by detailed simulations as well 
%as direct detailed observations at high redshift.  Theoretical progress in this area relies 
%mainly on phenomenological ``semi-analytic'' models (see e.g. Croton et al. \citealp{croton}, 
%Bower et al. \citealp{bower}, Monaco et al. \citealp{monaco}), which have had 
%success in reproducing locally-observed scaling relations, but are still heavily dependent on 
%simplified recipes for the physical mechanisms that drive galaxy evolution.  
%Direct observations of mass assembly at high redshift are therefore a key ingredient in probing the 
%role of various mechanisms in moderating galaxy formation. 

In recent years our observational knowledge of the high redshift Universe has increased enormously. 
A variety of selection techniques have been developed  
to create well-defined samples of z=1-6 galaxies, from 
UV to sub-millimeter wavelengths (e.g. Giavalisco et al. \citealp{giavalisco}, 
Steidel et al. \citealp{steidel}, Adelberger et al. \citealp{adelberger}, 
Franx et al. \citealp{franx}, Chapman et al. \citealp{chapman}, Daddi et al. \citealp{daddib}). 
These complementary and partially overlapping samples are now 
covering the critical redshift range around $z\sim2$, which is the epoch of major star formation 
and mass assembly in galaxies (Hopkins \& Beacom \citealp{hopkins}). 
Most of the first studies of these high-$z$ populations came from photometric 
measurements and comparisons with template spectral energy distributions (SEDs).  As a 
consequence, little was known about the detailed properties and dynamics of these systems, 
since such information can best be obtained through sensitive, high-resolution  
near-IR (rest frame optical) spectroscopy.
In the last few years, using increasingly sensitive IR detectors coupled with high throughput 
spectrometers and integral field techniques, it has become possible   
to carry out detailed imaging spectroscopy of the strongest rest-frame optical emission lines, 
allowing the assembly stage of massive galaxies to be studied directly through observations  
of the dynamical and detailed physical properties 
of the different populations at $z=1.5-3$ (see e.g. Erb et al. \citealp{erba}, \citealp{erbc}).
Indeed, the dynamical state of distant galaxies is the key property that enables a 
quantitative assessment of the main drivers of mass assembly and evolutionary processes.

In this context we have initiated \textit{SINS}, the high-z galaxy 
\textit{S}pectroscopic \textit{I}maging survey in the \textit{N}ear-IR with  
\textit{S}INFONI (F\"orster Schreiber et al. \citealp{survey}), making use of the 
capabilities of spatially resolved integral field spectroscopy to obtain new insights on the internal 
dynamics, sizes and morphology, dynamical masses, metal abundances and stellar populations of 
high-z massive star-forming galaxies.  With the direct measurement of these quantities, 
the first study of detailed dynamical properties of these sources has been successfully 
carried out.  We find evidence for large, massive, gas-rich rotating disks in place at 
$z\sim2$, requiring rapid and efficient mass accretion (F\"orster 
Schreiber et al. \citealp{natascha}; Genzel et al. \citealp{genzel}, \citealp{genzel08}; Shapiro et al. 
\citealp{kristen}; see also e.g. Wright et al. \citealp{wright}, Bournaud et al. 
\citealp{bournaud}, van Starkenburg et al. \citealp{vanstark}). 
While major ($<3:1$) mergers may form such disks in some 
cases (Robertson et al. \citealp{robertson06}, Robertson \& Bullock \citealp{robertson08}), 
on average a more continuous and steady mass accretion mechanism, 
such as through cold flows along cosmic web filaments (Keres et al. \citealp{keres}, Dekel \& 
Birnboim \citealp{dekel}, Dekel et al. \citealp{dekel08}, Sommer-Larsen et al. \citealp{sl03}), 
could be invoked. This leads to high star formation rates with longer time scales than the short 
bursts driven by mergers, and possibly induces the formation of central 
bulges via internal/secular dynamical processes (e.g. Noguchi \citealp{noguchi}, Immeli et al. 
\citealp{immeli}, Bournaud et al. \citealp{bournaud}).
The latest generation of cosmological semi-analytical and hydrodynamical simulations in 
the $\Lambda$-CDM paradigm indicate that minor mergers and/or smooth accretion are likely to 
dominate the assembly of the high-z populations, while major mergers would play a role 
in the smaller fraction of hyperluminous sources (see e.g. Kitzbichler \& White \citealp{kitzbichler}, 
Naab et al. \citealp{naab}, Guo \& White \citealp{guo}, Genel et al. \citealp{shy}).
A comparison of the dynamical properties of different samples of high-z galaxies 
observed in SINS (Bouch\'e et al. \citealp{nicolas}), shows that optically and IR-selected galaxies 
up to $z=2.2$ follow a similar velocity-size relation as $z=0$ late-type spirals, in strong 
contrast with sub-mm selected galaxies (see Tacconi et al. \citealp{tacconi}).

The spatially resolved integral field kinematics of high-z galaxies has therefore 
proven to be a powerful tool to study the major processes connected to the formation 
and evolution of galaxies in the early universe (see also Wright et al. \citealp{wright}, 
Law et al. \citealp{law}, Bournaud et al. \citealp{bournaud08} and van Starkenburg et 
al. \citealp{vanstark}). A robust measure of the 
internal kinematics of high-z galaxies is therefore a fundamental step in our understanding of 
the formation and early evolution of galaxies.

In this paper we present quantitative kinematic modeling of a sample of 
galaxies at $z\sim2$, selected from the SINS sample to be the most disk-like according to their
kinematics, showing a regular velocity ``spider-pattern'' with a clear velocity gradient and the
velocity dispersion peaking in the center. Thanks to the full two-dimensional coverage of our 
SINS datasets, we are able to model 
the internal dynamics of the warm gas as traced by the H$\alpha$ emission, deriving 
important constraints on the structure of these objects and on the evolution of 
their scaling properties, such as the Tully-Fisher relation (Tully \& Fisher \citealp{TF}). 

In Section~\ref{observ} we describe the sample selection, observations and data 
reduction, along with the extraction of the kinematic information from the 
observed datacubes. In Section~\ref{modeling} the disk model used and the fitting 
technique are presented, while the results are discussed in Section~\ref{results}. 
In Section~\ref{TFR} the stellar mass Tully Fisher relation at $z\sim2$ is presented, 
and our conclusions follow in Section~\ref{conclusions}. Throughout the paper we 
assume a $\Lambda$-dominated cosmology with $H_0=70\ \textrm{km} \textrm{s}^{-1} 
\textrm{Mpc}^{-1}$, $\Omega_m=0.3$ and $\Omega_{\Lambda}=0.7$. 
In this cosmology, $1\arcsec$ corresponds to $\sim 8.2$ kpc at $z=2.2$. 
All magnitudes are given in the Vega photometric system.

%__________________________________________________________________

\section{Sample and observations} \label{observ}

In the framework of the SINS survey (see F\"orster Schreiber et al. \citealp{natascha}, 
\citealp{survey}; Genzel et al. \citealp{genzel}, \citealp{genzel08};  
Bouch\'e et al. \citealp{nicolas}; Shapiro et al. \citealp{kristen}), we used the SINFONI 
near-IR integral field spectrograph on the ESO VLT (Eisenhauer et al. \citealp{frank}, 
Bonnet et al. \citealp{bonnet}) to measure the dynamics of the warm 
gas in a substantial sample of over 60 redshift $1.5-3.5$ massive star-forming galaxies, 
as traced by their H$\alpha$ emission redshifted into the $H$-band ($z \sim 1.5$) and $K$-band 
($z \sim 2.2$), or by [OIII] at $z>2.8$. 

\subsection{The SINS high-z disks sample} \label{samplesel}

F\"orster Schreiber et al. \citealp{survey} describe in detail the selection criteria of the SINS 
$z\sim2$ sample. Briefly, the SINS galaxies were selected to be massive star forming galaxies using 
a variety of complementary techniques in UV through IR wavebands (BX/BM criterion, Adelberger et al. 
\citealp{adelberger}; s-BzK, Daddi et al. \citealp{daddib}; K magnitude, Cimatti et al. 
\citealp{cimatti}, IRAC flux, Kurk et al. \citealp{kurk}).  
These selection criteria sample fairly luminous ($L_{Bol}\sim10^{11-12} L_{\odot}$) and 
massive galaxies ($M\sim10^{9-11.3} M_{\odot}$) with star formation rates of 10-300 $M_{\odot}$/yr.  
These represent the bulk of the cosmic star formation activity and stellar mass density at these redshifts 
(Reddy et al. \citealp{reddy}; Rudnick et al. \citealp{rudnick}; Grazian et al. \citealp{grazian}, 
Caputi et al. \citealp{caputi}).   
%As such, our sample 
%probes the early evolution of massive star-forming galaxies at high-$z$ (
%A further selection was based on the target visibility, H$\alpha$ line flux and night-sky 
%lines avoidance for H$\alpha$ at the redshift of the source. 
Here we present the dynamical modeling performed on a subsample of 18 galaxies 
(see Table~\ref{sampletab}, Fig.~\ref{exfit}), that are selected for showing ordered rotational 
signatures, i.e. a regular velocity pattern with a clear velocity gradient and a velocity dispersion 
that peaks in the center, as well as for their high S/N observations. 

In order to distinguish between rotationally dominated galaxies and sources with 
merger induced complex dynamics, we used, where possible, the method described in Shapiro et al. 
(\citealp{kristen}). Using kinemetry (Krajnovi\'c et al. \citealp{kinemetry}), we were able to 
quantify asymmetries in both the velocity and velocity dispersion maps of the 
warm gas, in order to empirically differentiate 
between major merging and non-merging (or minor merging) systems at high redshift. 
These criteria take full advantage 
of the wealth of information available with integral field data, which - unlike broad-band morphology - 
give direct information on the dynamical state of the galaxies.
The method was first applied to a subsample of galaxies with the best quality data 
(see Shapiro et al. \citealp{kristen}), and it has been now extended to a a larger subset of SINS 
galaxies (Shapiro et al. in preparation). In total 11 out of 18 galaxies of the sample 
discussed here were classified as rotationally dominated by kinemetry. For the remaining 7 sources  
(Q1623-BX663, Q2346-BX416, %K20-ID5, 
D3a-7144, D3a-4751, ZC1101592, GK2471, GK1084), we were not able to derive
    a quantitative classification due to the lower S/N of the data.
    We include them in this study since they closely resemble at a visual inspection 
the ones classified as disks in their dynamical properties (regular velocity pattern with 
clear velocity gradient, velocity dispersion peaking in the center), with no evidence of more 
complex merger induced dynamics.

The disk like systems selected for this study include 7 BX/BM galaxies 
selected by their rest-frame UV colors down to $R<25.5$, taken from the H$\alpha$ 
survey of Erb et al. (\citealp{erbc}), 2 galaxies selected with IRAC/Spitzer 
$m_{AB}(4.5 \mu m) < 23$ in the framework of the GMASS survey (Kurk et al. \citealp{kurk}), 
and 10 rest frame optical selected galaxies meeting the s-BzK color criterion 
introduced by Daddi et al. (\citealp{daddib}). These are taken from the K20 survey  
(Daddi et al. \citealp{daddia}), Deep-3a (Kong et al. \citealp{kong}) and 
z-Cosmos (Lilly et al \citealp{lilly}) surveys.
%2 galaxies selected in the GDDS survey with $Ks\leq 20.5$ and $I\leq 24.5$ 
%by Abraham et al. (\citealp{abraham}). 
%The galaxies in the sample selected for the modeling span the redshift range 
%from $z=1.4$ to $z=2.4$, and therefore their H$\alpha$ emission is redshifted 
%in the $H$ or $K$ band.

\subsection{Observations and data reduction}

We observed the selected galaxies with SINFONI (Eisenhauer et al. \citealp{frank}) 
at the VLT UT4 telescope during several campaigns between July 2004 and 
November 2007 as part of guaranteed time observations. 
The H or K band grating was used to sample the H$\alpha$ emission, depending 
on the redshift of the galaxy. Most of the observations were carried out in 
seeing limited mode, using the $0.250\arcsec \times 0.125\arcsec$ pixel scale, which provides a 
total field of view of $8\arcsec \times 8\arcsec$. The average resolution 
obtained for our 18 galaxies, as sampled by PSF reference star observations before and after 
every hour of target integration, is $\sim 0.5\arcsec$.
The spectral resolution provided is about 80 km/s Full Width Half Maximum (FWHM) in the $K$ band and 
100 km/s in the $H$ band.
For one source, D3a-15504, the presence of a suitable nearby reference star allowed 
us to use the correction provided by Natural Guide Star (NGS) Adaptive Optics 
(AO) observations. For two more objects, ZC782941 and Q2346-BX482, the Laser Guide 
Star (LGS) system PARSEC (Rabien et al. \citealp{rabien}, Bonaccini et al. \citealp{bonaccini}) 
was used to provide an average resolution of $0.2\arcsec$ (corresponding to only 1.25 kpc at $z=2$). 
In the AO observations of these three sources the $0.100\arcsec \times 0.050\arcsec$ pixel scale 
was used, providing a field of view of $3.2\arcsec \times 3.2 \arcsec$.
The observations and the data reduction were performed as described in 
F\"orster Schreiber et al. (\citealp{natascha}), Abuter et al. (\citealp{abuter}), 
and Davies et al. (\citealp{ric}).

The final result of the data reduction procedures is a flux-calibrated datacube, 
containing the full image of the galaxy observed at each wavelength. 
We are thus able to extract from the cube both the one-dimensional 
spectrum at each pixel (or integrated over a larger region) and monochromatic 
images of the field of view at different wavelengths. 
%The integral field technique is 
%therefore an ideal tool to study the dynamics of our high-z galaxies, providing 
%a more complete view than the long-slit based studies previously available at these 
%redshifts. 

\subsection{Kinematic extraction} \label{kinext}

To increase the S/N ratio, we median filtered the reduced cubes spatially using a 
FWHM=3 pixels, obtaining an effective resolution of $\sim 0.6\arcsec$, or 
$\sim 0.2\arcsec$ for the galaxies observed with the $0.05\arcsec$ pixel scale 
and AO. We then extracted from the smoothed cubes the continuum maps, line emission, 
velocity and velocity dispersion fields of H$\alpha$.
This was done by fitting a function to the continuum-subtracted spectral 
profile at each spatial position in the datacube, masking out the wavelength range 
contaminated with the [NII] line emission.
The function fitted was a convolution of a Gaussian with a spectrally
unresolved emission line profile of a suitable sky 
line, i.e. a high signal and non-blended line in the appropriate band. 
A minimization was performed in which the parameters of the Gaussian
were adjusted until the convolved profile best matched the data.
During the minimization, pixels in the data that consistently
deviated more than $3 \sigma$ from the average were rejected, and were not used in the analysis. 
The uncertainties on the measured velocity and velocity dispersion maps are evaluated through Monte 
Carlo realizations, perturbing the input data assuming Gaussian uncertainties. 
All details and issues of the kinematic extraction with the \textsf{IDL} routine
\textsf{LINEFIT} are described in more details in F\"orster Schreiber et al. 
(\citealp{survey}) and Davies et al. (in preparation).

%______________________________________________________________

\section{Dynamical modeling of the galaxies} \label{modeling}

The observed H$\alpha$ line emission kinematics of the selected galaxies 
can now be studied to robustly quantify their dynamical properties. 
This represents a major step forward in the modeling of the dynamics of high-z 
galaxies, as the full two-dimensional mapping from SINFONI is more complete and not biased by 
a priori assumptions about the kinematic major axis and inclination of the system 
as in long-slit spectroscopy (see also Wright et al. \citealp{wright}, Law et al. 
\citealp{law}). 

%We used for 18 galaxies the H$\alpha$ line emission, while for K20-ID5 we used the 
%[OIII] line data, redshifted to the H band, both for the better quality and to avoid 
%the strong contamination by broad H$\alpha$ emission from a central AGN.
We compare the velocity and velocity dispersion maps, derived from 
the SINFONI data as described in Sec.~\ref{observ}, with an exponential disk model
(scale height $h_z=0.1\arcsec$). Although a simple exponential disk could be an oversimplification 
of the structure in some of these high-z sources, this represents the best compromise 
between providing homogeneous reliable properties and avoiding over-interpretation 
of the data. A more detailed analysis is possible for the higher resolution and S/N 
datasets, using more complex mass distributions (see Genzel et al. \citealp{genzel08}). 

The observed intensity distribution of the H$\alpha$ emission line traces the 
location of the star forming regions across the galaxy, and has a clumpy 
and irregular distribution even for kinematically very regular disks. Therefore the line emission 
light distribution is used in the following only to evaluate the 
size of the emitting region. The best fitting disk parameters were derived 
using an optimized $\chi^2$ minimization routine, comparing the disk model 
with the observed velocity and velocity dispersion extracted as described in Sec.~\ref{kinext}.

\subsection{The disk model} \label{diskmod}

We create the disk models using %version 2.09 of 
the \textsf{IDL} code \textsf{DYSMAL}.
The code uses a set of input parameters to derive a datacube with two spatial and one 
spectral (velocity) axis, from which it is possible to extract 2D morphological and 
kinematical maps.
It begins by creating a face-on model of an axisymmetric galaxy with 3 spatial dimensions 
$[X_0,Y_0,Z_0]$, in which $Z_0$ is normal to the galaxy plane. 
The radial mass profile is specified by the user and can be based on exponential, 
Gaussian, Moffat, or power-law functions. The radial luminosity profile can be specified 
independently, since it can refer to different tracers that do not necessarily follow 
the mass distribution. The scale height of the disk must also be provided.
The 3D model is then rotated to the required inclination and position angles, to create 
a cube that has axes $[X_s,Y_s,Z_s]$ where $[X_s,Y_s]$ is the projection on the sky.
Although a warped disk can be modelled by specifying these 2 parameters as functions of 
radius, in all cases here the disk was modelled in a single plane.
To make this coordinate transformation more efficient, it is first necessary to 
identify those pixels in the resulting $[X_s,Y_s,Z_s]$ cube containing flux, and only 
fluxes for these pixels are calculated. 
The flux in each pixel is interpolated based on the sub-pixel location in the original 
$[X_0,Y_0,Z_0]$ cube. A second matching cube is created in which each pixel is assigned 
an appropriate line-of-sight velocity. The velocities are interpolated from a rotation 
curve that is itself generated from the combined mass profile (including a black hole mass 
if one is specified) and scaled to the total mass at a given radius.
The next step is, for each pixel $[X_s,Y_s,Z_s]$, to generate a Gaussian profile to model 
the line emission, whose scaling depends on the flux, whose center depends on the line-of-sight 
velocity, and whose width depends on both the instrumental broadening specified by the user and 
a local isotropic velocity dispersion depending on the z-scale height. 
This z-velocity dispersion $\sigma_{01}$ is estimated in one of two ways:
\begin{enumerate}
	\item using an approximation appropriate for large thin disks, in which it depends 
		on the scale height and mass surface density (Binney \& Tremaine \citealp{binney}, eq. 4.302c):
		\begin{equation}
			\sigma_{01} \sim \sqrt{\frac{v^2(R)\ h_z}{R}}
			%\frac{h_z}{R}=0.5 \left(\frac{\sigma_{01}}{v(R)}\right)^2
		\end{equation}
		where $h_z$ is the scale height and $v(R)$ the rotational velocity at radius $R$;
	\item assuming that the vertical motion is restricted to the inner part of an extended 
		mass distribution and does not feel the entire gravity:
		\begin{equation}	\label{sigmazero}
			\sigma_{01}\sim \frac{v(R)\ h_z}{R}
			%\frac{v(R)}{\sigma_{01}} \sim \frac{R}{h_z}
		\end{equation}
		which is more suitable for a thick or compact disk (Genzel et al. \citealp{genzel08}), 
		and it was used in the modeling discussed below.
\end{enumerate}
The total z-velocity dispersion in the model is then given by $\sigma_0=\sqrt{\sigma_{01}^2+\sigma_{02}^2}$, 
where $\sigma_{02}$ is an additional component of isotropic velocity dispersion throughout 
the disk that can be added by the user. This term was introduced in \textsf{DYSMAL} 
to account for instrumental broadening. It can also be used to include 
any additional possible contribution to the observed velocity widths from, e.g., rapid mass 
accretion, strong feedback from star formation, or unresolved non-circular motions (hereafter 
globally called ``random motions'' for simplicity).
This yields a line-of-sight velocity profile for each pixel in $[X_s,Y_s,Z_s]$.
Then, at each $[X_s,Y_s]$ location, the velocity profiles for all the $[Z_s]$ positions 
are summed to yield the full line-of-sight velocity distribution at that point.
This creates a cube $[X_s,Y_s,V]$ with two spatial and one velocity axes.
The final step is to convolve each $[X_s,Y_s]$ plane in this cube with the spatial beam and to rebin 
to the desired pixel scale.

The reason for beginning with a galaxy in 3 spatial dimensions is to ensure that there 
is always enough spatial resolution along the projected minor axis to yield a reliable 
line-of-sight velocity distribution. This can be an issue when the inclination is high, 
and in such cases a direct 2D projection is insufficient.
It also enables one to apply a mask that modulates the flux distribution, for example to 
simulate a knotty or clumpy structure. The code allows one either to generate a mask 
containing random knots; or to input one if, for example, a more ordered pattern such 
as spiral arms is required. In either case, the mask is applied to the galaxy while it 
is face-on, in the $[X_0,Y_0]$ plane. For this high-z study, however, a plain exponential disk 
was used to keep the number of free parameters as low as possible.

The velocity dispersion across the galaxy disk is an important parameter to consider.  
When there are significant random motions, as implied by the large observed dispersion, they 
should also support some of the mass. This is currently not implemented in DYSMAL, 
which assumes that all the mass is supported by ordered rotation in the disk plane.
Thus, if $v/\sigma_0 \lesssim \sqrt{3}$, the total system mass may be underestimated by a factor 
of 2 or more. The average $V/\sigma_0$ for our sample is $\sim4.4$, producing an effect of $\sim 12\%$.

We account for spatial beam smearing from the PSF and velocity broadening  from instrumental 
spectral resolution in our modeling, as we compare this spatially and spectrally convolved model 
disk to the observations. Hence, the best-fit quantities represent the intrinsic 
properties derived by properly taking into account these observational effects.

\subsection{Fitting technique: an evolutionary approach} \label{fitting}

Given the large number of parameters and the potential for many local minima in 
the likelihood/$\chi^2$ space, it is important to use a reliable fitting algorithm, 
i.e. robust and efficient one to find the global extremum. We selected for this task 
genetic algorithms, which are heuristic search techniques that 
provide a means to efficiently sample the parameter space, reproducing in a computational 
setting the biological 
effect of evolution by natural selection (e.g. Holland, \citealp{holland}). These
kinds of algorithms have the advantage of being robust and efficient  
in finding the global maximum of a given function even in a very complex parameter 
space, with a reasonable number of iterations and without the need of a good set 
of initial guesses. It thus represents an ideal tool for our specific problem.

We use a modified version of \textrm{PIKAIA}, a genetic algorithm-based 
optimization routine by Charbonneau (\citealp{charbonneau}). The algorithm starts with a 
random population of ``individuals'' that represent a single location in parameter space 
with a complete set of the parameters we want to fit. 
The fitness of each individual is evaluated through their $\chi^2$, and the 
probability to survive in the next iteration is set to be proportional to this 
value. In this way the more suitable parameter sets are selected, but at the same 
time the parameter space is efficiently sampled by ``breeding'' between different 
individuals and allowing random ``mutations'' and ``crossovers'' between the 
parameters.  
Therefore, in each iteration the whole population becomes more and more suitable, 
while still sampling the parameter space. When the requested tolerance is reached, 
or at the end of a given number of iterations, the best fitting ``individual'' 
is chosen as the solution.

The genetic algorithm allows us to efficiently minimize the differences between
the exponential disk model and the observed velocity and velocity dispersion 
fields of the H$\alpha$ emission line. 
Simulations on test galaxies have shown that leaving the rotation center as a 
free parameters in the fitting significantly worsens the the final results, even 
after increasing the number of maximum iterations allowed (see also Shapiro et al. 
\citealp{kristen}). We therefore locate the center
using the continuum light distribution of the galaxies, which corresponds roughly 
to rest frame $R$ band as observed in the near-IR at $z\sim2$.  
In the SINFONI data of our galaxies, the continuum is detected, although typically at a lower 
S/N than the emission lines.
The resulting intensity map is generally smoother and distinct than 
the emission line (F\"orster Schreiber et al. \citealp{natascha}, Genzel et al. 
\citealp{genzel08}, Shapiro et al. \citealp{kristen}), as it is also the case 
for local disk galaxies (e.g. Daigle et al. \citealp{daigle}). These continuum intensity maps 
show a clear peak in the central region, especially in the BzK selected galaxies, and represent 
a better tracer of the mass distribution (dominated by stars) in the galaxies than 
the more clumpy line emission (tracing active sites of star formation). 
The center is identified by 
computing the intensity weighted mean of the light profile of the central pixels
along both minor and major axes, and it is kept fixed in the disk fitting procedure.

\begin{figure}
	\centering
	{\includegraphics[width=10cm,keepaspectratio]{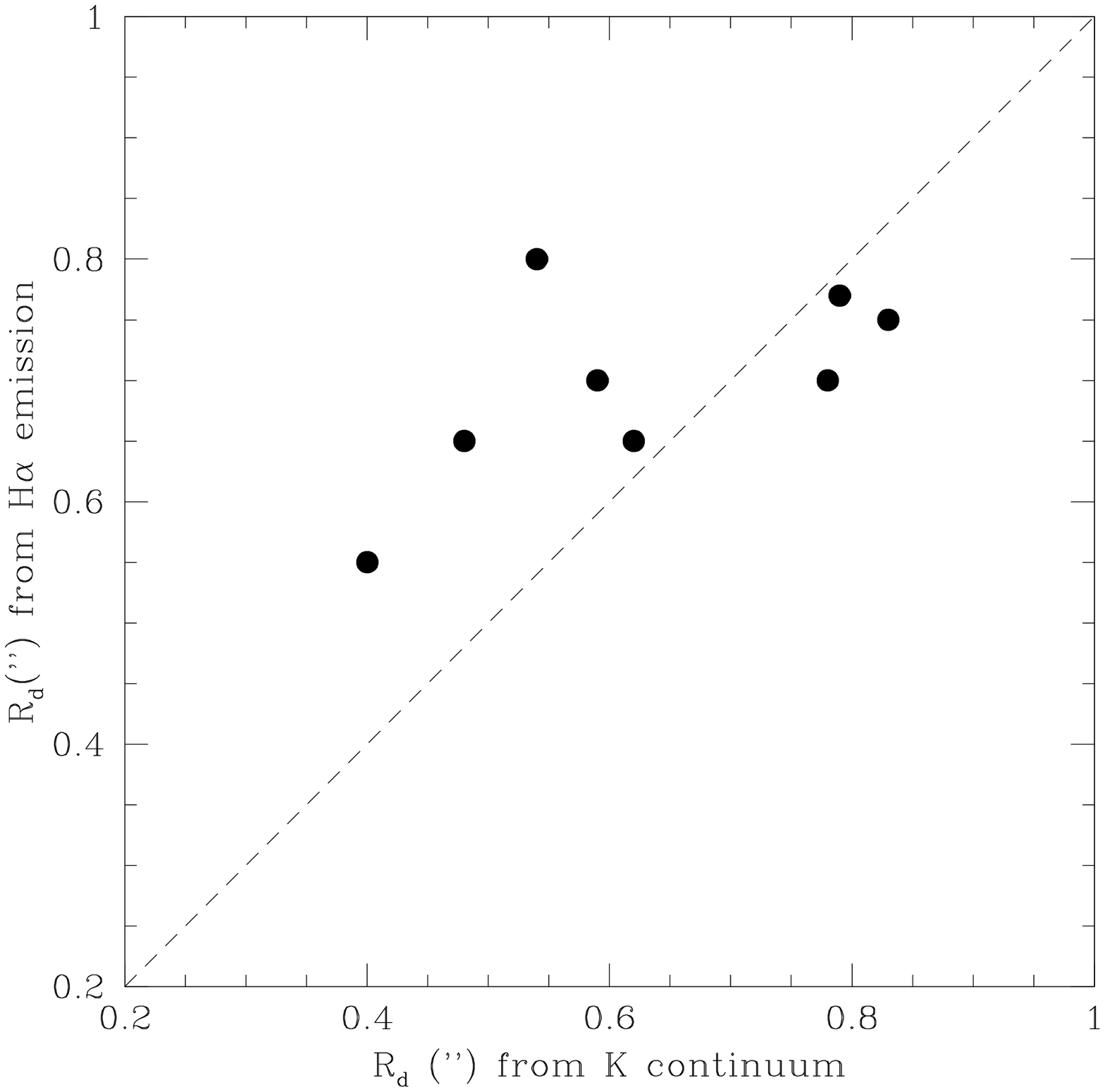}}
	\caption{Comparison between the scale length $R_d$ measured from the line emission 
	in our SINFONI data with the value obtained by fitting a Sersic profile to 
	HST and ISAAC K-band continuum images 
	available for a subsample of galaxies (5 BX, 2 K20 and 1 GMASS). 
	%The point deviating from the correlation between the 
	%two measurament is K20-ID5, where the K-band continuum is probably strongly contaminated by 
	%the point-like AGN.
	}
	\label{rdfig}
\end{figure}

The other parameter that is provided as an input to the fitting routine is the 
scale length $R_d$ of the disk. This parameter is in fact  
better constrained by the emission intensity than by the dynamical properties 
of the gas. In this case we used the higher S/N line emission 
image of the galaxies to quantify it.  The disk H$\alpha$  
distribution has been found to follow the continuum for local spiral galaxies, with consistent derived 
scale lengths (e.g. Tacconi et al. \citealp{tacconi86}, Koopmann et al. \citealp{koopmann}, 
Hanish et al. \citealp{hanish}).
%We tested two different approaches to measure the scale length of the disks 
%from the line emission map of the galaxies. First we used a curve of growth of the 
%photometry, obtained in concentric ellipses around the center, to determine the half 
%light radius $R_{1/2}$ of the observed emission. 
%Assuming an exponential disk profile, the scale length is given by 
%$R_{1/2}/1.6783$. However the half light radius evaluated in this way 
%represent a lower limit to the actual one, as the external regions 
%of the galaxies, where the  the surface brightness is much lower, are dominated by 
%the noise. However, due to the very different observing 
%conditions, integration time and data quality for the galaxies of the sample, it 
%is not easy to correct for this effect.
From an analysis of different SINS and simulated galaxies, we found that the most robust way
to measure the scale length of the disks given the S/N and resolution of the data is to fit 
a linear Gaussian profile along the major axis of the line intensity maps. This 
method has the advantage of using just the bulk of the brighter line emission, avoiding 
the problems related to the surface brightness at the edges of the galaxies.
Following Bouch\'e et al. (\citealp{nicolas}), 
we assumed that the derived HWHM, once corrected for the observed seeing, corresponds to 
the scale length $R_d$ of the disk. This assumption is supported by model disk fits, 
incorporating the beam smearing due to the $\sim0.5\arcsec$ FWHM resolution of our data. 
These simulations show that $R_d$ measured in this way is likely to be overestimated 
by no more than $\lesssim 15\%$. 
Using available HST/NICMOS continuum imaging (see F\"orster 
Schreiber et al. in preparation) and deep ISAAC imaging in GOODS (data release v1.5:
http://www.eso.org/science/goods/releases/20050930/) of a subsample of 
these sources (1 GMASS, 2 K20 and 5 BX), we find reasonable agreement between our measured 
scale length and the ones derived by fitting an exponential profile to these K band images 
with GALFIT (Peng et al. \citealp{peng}). The results are shown in Fig.~\ref{rdfig}. 

According to simulations, a choice of a scale length $R_d\ 15\%$ smaller would provide a 
dynamical mass $\sim10\%$ lower, and a $\sim30\%$ smaller one a mass $\sim20\%$ lower. 
We note that the value obtained for the maximum rotational velocity $V_{max}$ is instead 
much less sensitive to a variation of the scale length, as the fitting routine has to reproduce 
the observed velocity pattern.
%A $\lesssim10\%$ overestimation of the total dynamical mass due to the $R_d$ determination has 
%to be compared with a possible underestimation due to the large z-scale heights $h_z$ inferred for 
%the $z\sim2$ galaxies (see F\"orster Schreiber et al. \citealp{natascha}, Genzel et al. 
%\citealp{genzel08}), for which the thin disk approximation holds at the $\sim5-15\%$ level of 
%velocity estimation (see Sect. \ref{diskmod} and Binney \& Tremaine \citealp{binney}, 
%chapter 2). We therefore assume a global systematic uncertainty of $10-15\%$,
%which is sufficient for our modeling. 
However, we also note that if a flatter mass distribution is a more appropriate description of 
the structure
of these high-z disks, e.g. the ring-like structures discussed in Genzel et al. (\citealp{genzel08}), 
the true enclosed masses would then be up to $\sim10-30\%$ lower than the one obtained with the 
method presented here.

The inclination of the disk with respect to the line of sight $i$, the position 
angle on the plane of the sky $\theta$, the total enclosed mass $M_{dyn}$, the velocity zero 
point $v_0$ at the rotation center, and the constant global isotropic dispersion term $\sigma_{02}$  
(see eq.~\ref{sigmazero}), are left as free parameters in the minimization. 
The model disk obtained is then rescaled as it would appear at the redshift of 
the source, re-binned to the pixel size of the current observation, and smeared 
using the FWHM of the PSF observed (see Sect.~\ref{observ}). 
In every iteration of the genetic algorithm, the $\chi^2$ is evaluated for each 
``individual'' from the squared difference between the observed velocity and
velocity dispersion maps and the modeled ones. The difference in each pixel is 
weighted using the uncertainties on the velocity and on $\sigma$ derived 
as described in Sec.~\ref{observ} at that location, e.g.:
\begin{equation} \label{chi}
	\chi^2(vel)=\sum_{i} \frac{(V(obs)_i-V(mod)_i)^2}{\Delta V(obs)_i^2} 
		\times \frac{1}{N_{pix}-N_{param}}
\end{equation}
where $N_{pix}$ is the number of pixels with S/N over the threshold (see Sect.~\ref{kinext}) 
used in the analysis. The best fitting parameter set is therefore taken as the solution. The 
uncertainties are evaluated following Avni (\citealp{avni}) method to 
calculate confidence limits for a multiple number of parameters, 
studying the variation of $\chi^2$ around the optimized minimum. We assumed as $90\%$ 
confidence interval where the reduced $\chi^2$ is increased by $\Delta \chi^2=6.25$ 
for the three ``interesting parameters'' (inclination, dynamical mass and $\sigma_{02}$) 
used in the analysis.

%______________________________________________________________

\section{Results} \label{results}

For each galaxy in the sample we show, in Fig.~\ref{exfit}, the results of the best fit
with the exponential disk model, for both the velocity and velocity 
dispersion fields. The disk model is able to reproduce the 
global dynamical pattern of the galaxies, although in the observed maps small 
deviations from a perfect ``spider-diagram'' are observed. These deviations are 
clear as residuals in the difference image between the observed maps and the 
model, and in the $\chi^2$ maps, defined as above (eq.~\ref{chi}).
The total reduced $\chi^2$ obtained for the galaxies vary between 0.2 and 20, but 
comparing directly the different $\chi^2$ values between the 
galaxies is not a reliable criterion for evaluating the quality of the modeling for 
the different sources nor for distinguishing ordered motions from disturbed kinematics due to interactions. 
In fact the $\chi^2$ test is, as expected, sensitive to the average S/N of the 
observation, since the value is weighted using the uncertainties on the velocity 
and dispersion. Since the S/N of the data varies from source to source, 
the separation between rotating disks and disturbed kinematics is 
performed using the kinemetry criteria developed by Shapiro et al. (\citealp{kristen}, 
see section 2).

The results of the modeling are summarized in Tab.~\ref{fitresults}. 
The dynamical mass reported is the total mass in the best fitting exponential disk
at a radius of 10 kpc ($\sim 1.2\arcsec$ at $z\sim2$). This radius was chosen to 
roughly match the photometric radius used to measure the ``total'' fluxes of the galaxies 
in the different filters, which were used in the SED fitting to calculate the stellar 
masses (see following section), allowing a meaningful comparison of the two 
quantities.
$V_{max}$ is the maximum velocity in the exponential disk model. The main source of uncertainty is 
the determination of the inclination with respect to the plane of the sky.
Moreover, the stated errors on the dynamical mass do 
not include possible systematic effects, e.g. due to the value used for $R_d$ or due to the choice of 
the exponential mass profile (see Sect.~\ref{fitting}). 

The measurement of $\sigma_{02}$ is mainly constrained from the dispersion in the outer part of the disks,
where the effects due to the beam smearing of the steep central rotation curve are less prominent. 
Some of the galaxies were too compact, or with insufficient data quality to derive a robust measure 
of this quantity. In one case, Q2343-BX389, the irregular dispersion profile was also preventing a 
unique and robust estimate for the isotropic dispersion using this technique.
In agreement with the earlier findings of F\"orster Schreiber et al. (\citealp{natascha}) 
and the modeling of Genzel et al. (\citealp{genzel08}), the majority of the star forming 
galaxies modeled have a large component of local random motion, with $\sigma_{02}\sim 30-80$ 
and a average $V_{max}/\sigma_{02} = 4.4$, with a typical uncertainty of $\pm20$ km/s.
Low values of the V/sigma ratio in these galaxies suggest that a significant fraction of star-forming 
disks at $z \sim 2$ may be highly turbulent and geometrically thick. Concurring evidence from kinematics
has now also been found in other $z \sim 1.5-2.5$ disk-like systems (e.g., Wright et al. \citealp{wright}, 
van Starkenburg et al. \citealp{vanstark}) and is supported by the large thickness ($\sim 1$ kpc) inferred
from morphological analysis of candidate edge-on disks based on high resolution HST imaging 
(Elmegreen \& Elmegreen \citealp{elmeg}). 

The dynamical evidence of the presence of such large, rotating and turbulent disks is in agreement with 
the scenario in which rapid, smooth gas accretion from the halo may play a significant role in the the 
formation of massive galaxies (Genzel et al. \citealp{genzel08}). This picture is also 
supported by recent broad-band observations of high-$z$ populations, which have found a 
fairly tight relation between stellar mass and star formation rate, a sign that 
steady star formation rates, rather than merger induced short luminous bursts, dominate 
the stellar assembly of these galaxies (Noeske et al. \citealp{noeske}; Elbaz et al. 
\citealp{elbaz}; Daddi et al. \citealp{daddi07b}). The most recent dark matter simulations
such as the Millennium (Springel et al. \citealp{springel}) show that subsequent mergers are 
not frequent enough to convert all $z\sim2$  
disks into elliptical galaxies at $z=0$, and that most of the mass growth
is achieved via mergers less intense than 10:1 or smooth accretion 
(Genel et al. \citealp{shy}). Recent hydrodynamic simulations (see e.g. Sommer-Larsen et al. 
\citealp{sl03}, Ocvirk et al. \citealp{ocvirk}, Dekel et al. \citealp{dekel08}) also show that 
steady, cold gas streams, which penetrate effectively through the shock-heated media of dark-matter 
haloes, are a main accretion mechanism for galaxies in this redshift range.

\subsection{Comparison between dynamical and stellar mass} \label{stmass}

Thanks to the multiband coverage of the different fields from which we drew our SINS 
targets (see Sect.~\ref{samplesel}), we were able to perform a Spectral Energy 
Distribution (SED) analysis using the available broad-band photometry from the 
observed optical to near-IR/mid-IR. In particular, as described in detail in 
F\"orster Schreiber et al. (\citealp{survey}), we used data from published works by 
Erb et al. (\citealp{erbb}, BX), Daddi et al. (\citealp{daddia}, K20), 
Kong et al. (\citealp{kong}, D3a), Kurk et al. (\citealp{kurk}, GMASS), 
Capak et al. (\citealp{capak}, COSMOS), as well 
as photometry kindly provided to us by McCracken et al. (in preparation, COSMOS).
SED modeling has been carried out by these authors but using different model 
ingredients or assumptions. In order to ensure more uniformly derived properties 
among our sample, we re-modeled all the galaxies following the prescriptions 
described in F\"orster Schreiber et al. (\citealp{nataschased}, see also F\"orster 
Schreiber et al. \citealp{survey}). Our results do not differ significantly from the 
published ones.
We used the solar metallicity Bruzual \& Charlot (\citealp{bruzual}) stellar population 
synthesis models, with a Chabrier Initial Mass Function (IMF) between 
0.1 and 100 $M_{\odot}$, and the Calzetti et al. (\citealp{calzetti}) reddening law. 
We considered three combinations of star
formation history and dust content: constant star formation
rate (CSF) and dust, an exponentially declining star formation
rate with e-folding timescale of t = 300 Myr and dust
($\tau300$), and a dust-free single stellar population formed instantaneously. 
We adopted the best of those
three cases based on the reduced chi-squared value of the fits
(for all the galaxies in the sample, this is either CSF or $\tau300$).
The redshift is kept fixed to the spectroscopic redshift 
measured from the emission lines detected, while the stellar mass $M_*$, SFR, star formation 
history, stellar population age and visual extinction $A_V$ are free parameters. 
We evaluate the uncertainties through Monte Carlo simulations, where the input photometry 
is varied randomly assuming the photometric uncertainties are Gaussian. 
The fitting results for the galaxies of the sample are reported in Table~\ref{sampletab}.  

The reported stellar masses are affected by various uncertainties and systematic errors. 
In particular, different stellar population synthesis models do not reproduce a consistent picture 
of evolution in the rest frame NIR (see Marchesini et al. \citealp{marchesini}). 
For example, Maraston et al. (\citealp{maraston}) suggests that the Bruzual \& Charlot 
(\citealp{bruzual}) stellar population synthesis models used here may overestimate the stellar masses 
up to a factor of $\sim2$ when the bulk of the population in the age range between $\sim 0.5$ 
and $\sim 2$ Gyr. This is due to a different approach in the treatment of the thermally-pulsing 
asymptotic giant branch (TP-AGB) phase. However, findings from much larger samples, where the 
effects of SED modeling can be better and more reliably investigated on a statistical basis, suggest 
that the correction between the two different approaches can be negligible or up to $\sim30\%$
(see e.g. Berta et al. \citealp{sberta}, Marchesini et al. \citealp{marchesini}, Wuyts et al. \citealp{wuyts}). 
Accordingly, using the different library of synthetic spectra
of Charlot \& Bruzual (2007), the stellar masses estimated for our small sample change individually by 
no more than 20\%, while the median of the whole sample remains consistent within 2\%.
In any case, this correction would make the stellar masses lower, and the offset 
in the zero point of the high-z Tully-Fisher relation discussed in Sect.~\ref{TFR} will be 
correspondingly larger.

In four objects of the sample there are independent indications of the presence of an AGN: 
dynamical and UV spectral evidence suggests the presence of an AGN in D3a-15504 (see Genzel et al. 
\citealp{genzel}, Daddi et al. in preparation), BX663 and D3a-7144 show AGN features in optical spectroscopy
(Erb et al. \citealp{erbc} and Daddi et al. in preparation respectively), 
while for D3a-6004 a 24$\mu m$ excess was observed by Daddi et al. (in preparation), and a large 
[NII]/H$\alpha \sim 0.63$ ratio was measured in our data 
in the nuclear region of the galaxy (Buschkamp et al. in preparation). The inspection 
of the optical to K band available SED for these galaxies suggests that this AGN contribution may be biasing 
the stellar mass estimate towards higher stellar mass in the latter two cases (see F\"orster Schreiber 
et al. \citealp{survey}). We therefore conservatively consider the stellar mass estimate for D3a-6004 and 
D3a-7144 as an upper limit. 

\begin{figure}
	\centering
	{\includegraphics[width=10cm,keepaspectratio]{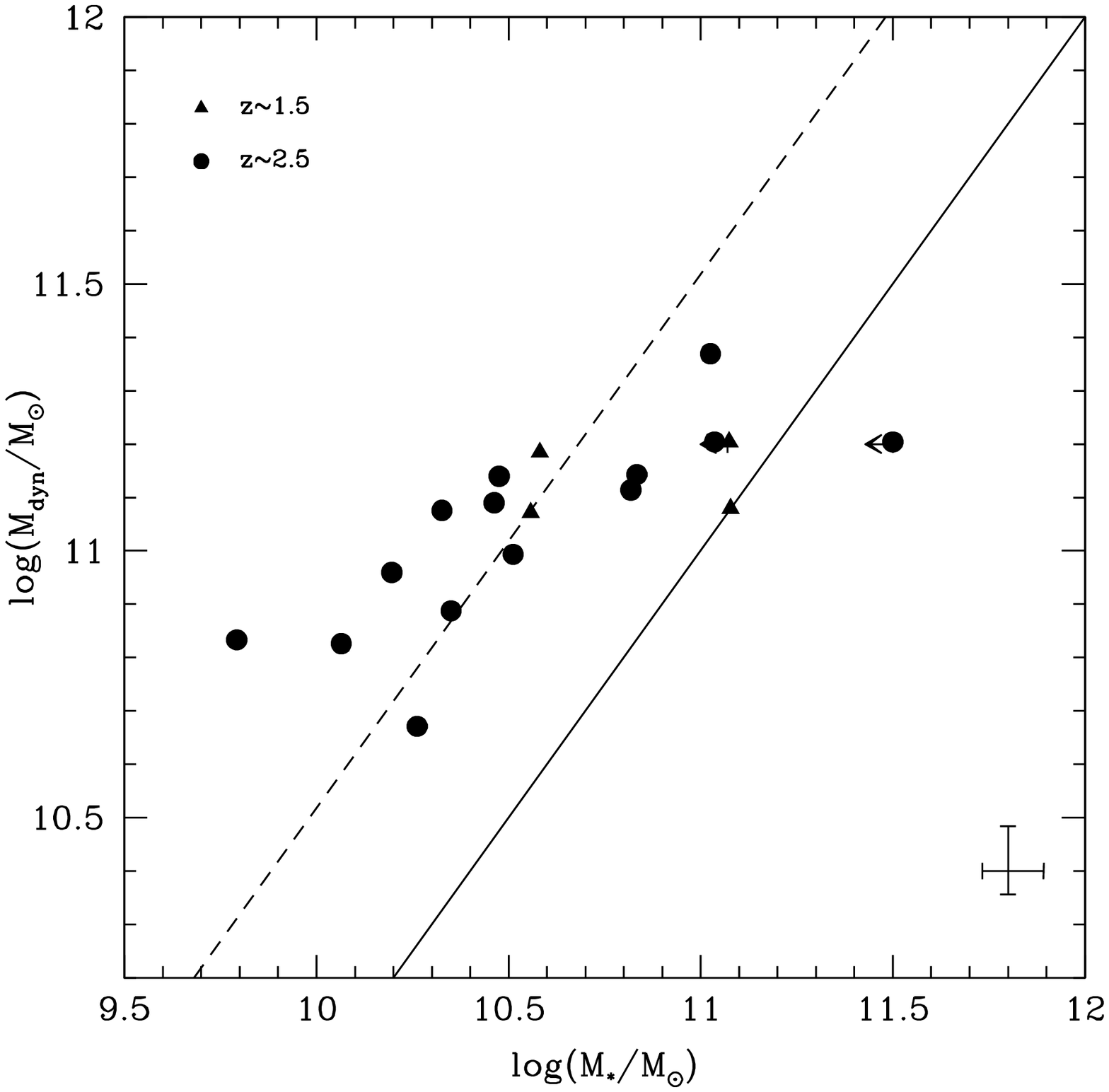}}
	\caption{Comparison between dynamical and stellar mass for the galaxies 
	of the sample with available multiband photometry. The dynamical mass is the total mass
	in the best fitting model up to $R=10$ kpc, while the 
	stellar mass is derived by SED modeling assuming a Chabrier IMF. 
	The filled triangles are the $z\sim1.5$ galaxies from our sample, while the 
	circles are at $z\sim2.2$. The solid line 
	indicates equal mass, while the dashed one indicates a dynamical mass 3.3 times 
	larger. The error bar in the lower right corner represents the average fitting uncertainties 
	in both directions. }
	\label{dynstel}
\end{figure}

The stellar masses derived for the sample are between $1-30 \times 10^{10}\ M_{\odot}$, 
with an average of $6.3\ 10^{10}\ M_{\odot}$
confirming that all the selected galaxies are drawn from the massive star forming galaxy 
population at high-z. %The average derived age is $0.5$ Gyr.
Thanks to the SED results and our dynamical model, we are able to compare the 
stellar and dynamical mass of the galaxies. In Fig.~\ref{dynstel} these are 
compared for galaxies in which multi-band photometry is available to constrain 
the SED fitting. We observe remarkable correlation between the two quantities: 
the Spearman correlation test gives a probability $P=2\cdot10^{-5}$ that the two masses 
are uncorrelated.

The resulting average dynamical mass at $R=10 kpc$ is $\sim 3.3$ times larger than 
the average stellar mass, excluding the sources with a possible AGN contribution. 
Therefore, given a dark matter contribution of $\sim 40\%$ at this radius to the 
total dynamical mass of such disks (see Genzel et al. \citealp{genzel08}), 
we infer a large fraction of gas in the sample galaxies ($M_{gas}\sim M_{*}$, or $\sim 30\%$), 
although with very large uncertainties. 
This estimate may be compared with a gas mass derived from the star formation rate surface density
using the Schmidt-Kennicutt relation (see Bouch\'e et al. \citealp{nicolas}). In this way we derive an 
average gas contribution of 23\% to the total dynamical mass of the galaxies, where the difference can be 
accounted for by the large uncertainties in both our estimate and the extinction correction needed to 
derive the star formation rates from the detected H$\alpha$ fluxes.
The gas fraction estimate is also in agreement with the findings of Daddi et al. 
(\citealp{daddi08}), who directly observed molecular gas in two $z\sim1.5$ 
BzK galaxies, deriving gas fraction 
as high as $M_{gas}\sim M_{*}$. 
An alternative possibility is that the stellar mass is significantly 
underestimated for most of the sample, presumably because of a faded old stellar population 
that is not detected in the multicolor SED, but would still contribute to the total dynamical 
and stellar mass (see the discussion by, e.g. Shapley et al. \citealp{shapley}, Papovich 
et al. \citealp{papovich}, and for the galaxies with available IRAC photometry 
Wuyts et al. \citealp{wuyts}), or that the 
dynamical mass is overestimated due to the systematic effects discussed in Sect.~\ref{fitting}.

\begin{figure}
	\centering
	{\includegraphics[width=10cm,keepaspectratio]{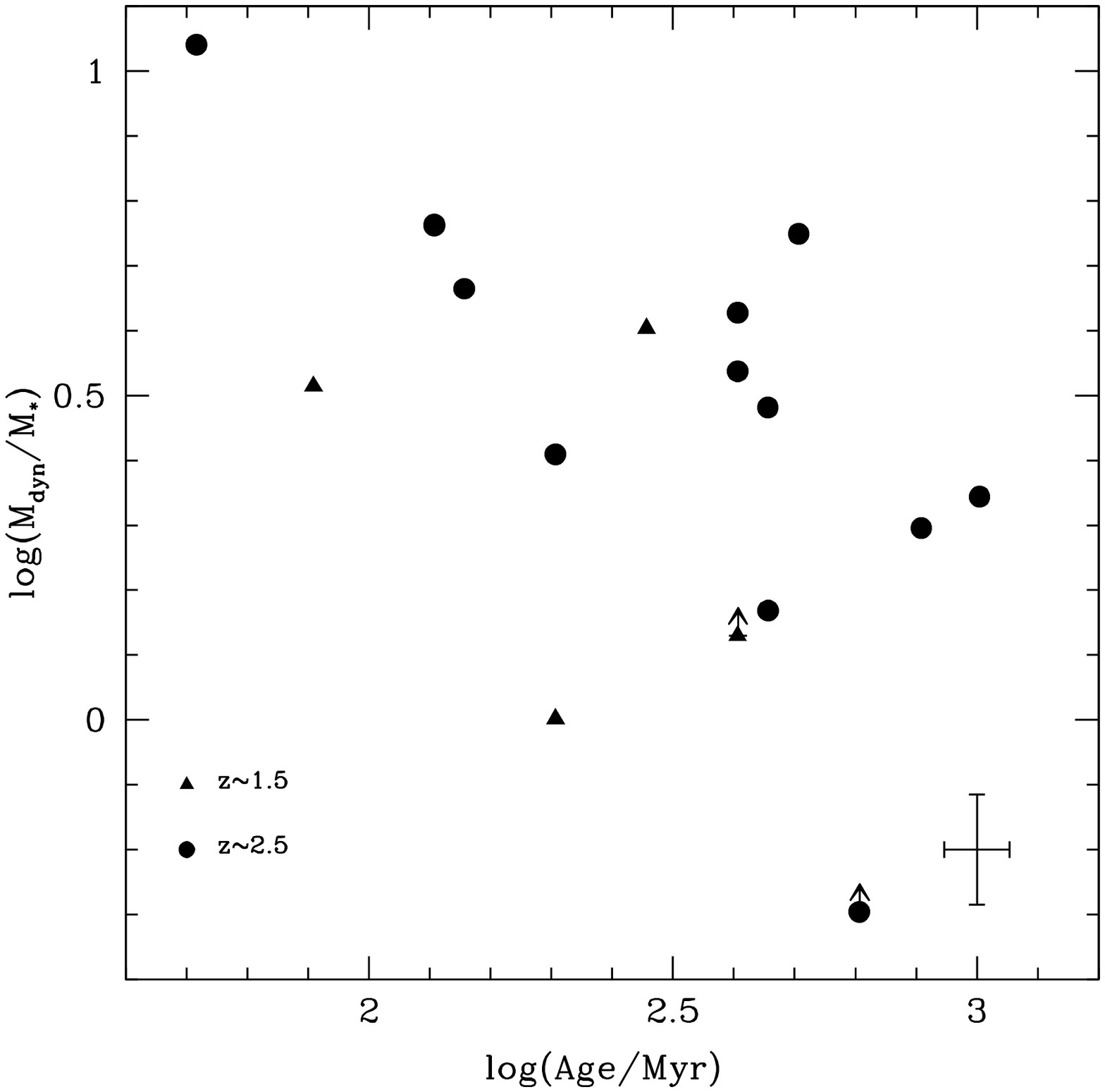}}
	\caption{The mass ratio $M_{dyn}/M_*$ is plotted against the best fitting 
	age from SED modeling. The filled triangles are the $z\sim1.5$ galaxies from 
	our sample, while the circles are at $z\sim2.2$. Galaxies with younger 
	stellar populations show a higher $M_{dyn}/M_*$, suggesting larger gas 
	fractions. However, the large observed scatter may suggest a significant 
	contribution from continuous gas accretion. The error bar in the lower right 
	corner represent the average fitting uncertainties in both directions.}
	\label{ages}
\end{figure}

Although the stellar population age is not strongly constrained by the SED fitting due to the 
uncertainty on the star formation history and the degeneracy with extinction, 
we note that the objects with the higher $M_{dyn}/M_*$ generally correspond to younger best 
fitting ages, similar to what was discussed by Erb et al. (\citealp{erbb}), and that none of the 
youngest galaxies has a small mass ratio (see Fig.~\ref{ages}). The significance of the 
correlation according to the Spearman test gives a probability $P=0.08$ that the two quantities are 
uncorrelated. The scatter is in fact high, probably suggesting that these galaxies are not closed systems, 
and that  an important contribution from continuous gas accretion must be considered. 
This again suggests that the galaxies in the sample are star forming disks with significant 
gas fraction and gas accretion, where the gas consumption correlates with the stellar population age.

%______________________________________________________________

\section{The $z\sim2.2$ Tully-Fisher relation} \label{TFR}

The relation between luminosity (or stellar mass) and maximum rotational 
velocity of disk galaxies has long been known (Tully \& Fisher 
\citealp{TF}). This Tully-Fisher relation (TFR) has been used as a distance 
indicator for disk galaxies (e.g. Tully \& Pierce \citealp{tullypierce}, Sandage 
\citealp{sandage}), but also as a key feature for understanding the structure and evolution of these 
galaxies, as it links directly the angular momentum of the dark halo with 
the luminosity (or mass) of the stellar population in the disk.
In fact, according to the standard model for disk galaxy formation (Fall \& Efstathiou 
\citealp{fall}), disks form out of gas cooling down from a hot halo associated with the dark 
matter potential well, maintaining its specific angular momentum and settling 
in a rotationally supported disk (Mo et al. \citealp{mo}). Consequently, the structure and dynamics 
of disk galaxies are expected to be closely correlated with the properties of the dark matter halo 
in which they are embedded, and the observable structural parameters of disks at different 
cosmic epochs can be used as tracers of the properties of the corresponding halos. 
The evolution of the stellar mass TFR is expected to be related both to the conversion of gas 
into stars and to the inside-out growth of the dark matter halo by accretion.  
In fact, while the extent of the halo grows considerably 
with time, the circular velocity of the halo grows less keeping the rotation curve approximately 
flat to larger and larger distances. The accretion of the dark matter is followed by accretion of 
baryonic gas, which is subsequently converted into stars by ordinary star-formation
in the disk. However, the details of the process and the amount of evolution expected 
depends strongly on the model assumptions, accretion mechanism adopted and the timescale 
needed to convert the gas into stars.
Any successful model of disk formation should then be able to reproduce the slope, 
zero-point, scatter and redshift evolution of the TFR and other scaling relations. 
However, reproducing simultaneously the TFR, the shape and normalization 
of the luminosity function, and the observed sizes, metallicities and colors of disk galaxies is 
still a challenging task for current models (Courteau et al. \citealp{courteau}).
Therefore, as the normalization, slope and evolution of the TFR depend strongly on the 
prescription used for the star formation and feedback, as well as on the disk formation mechanism and 
cosmological parameters, it represents a crucial test for galaxy evolution models 
(e.g. Silk et al. \citealp{silk}; Steinmetz \& Navarro \citealp{steinmetz}; van den Bosch 
\citealp{vanbosch}, \citealp{vanbosch02}; Sommer-Larsen et al. \citealp{sl03}; 
Dutton et al. \citealp{dutton}).

In recent years an increasing number of dynamical observations of disk 
galaxies in the local and intermediate redshift Universe have allowed us to begin
exploring the evolution of the TFR with redshift. 
There are still discrepant results on a possible evolution at 
intermediate redshifts of the tight relation observed in local samples of disk 
galaxies (e.g. Haynes et al. \citealp{haynes}, Pizagno et al. \citealp{pizagno07}) 
between the absolute magnitude and the maximum rotational velocity. For example, while 
Vogt et al. (\citealp{vogt}) reported very little evolution of the B-band TFR 
up to $z\sim1$, other groups (see e.g. Simard \& Pritchet \citealp{simardpr}, Barden et al. 
\citealp{barden}, Nakamura et al. \citealp{nakamura} and B\"ohm et al. \citealp{bohm}) 
found a strong brightening of $\sim1-2$ mag in B-band luminosity over the same 
redshift range. 

Furthermore, the interpretation of a possible evolution of a luminosity based 
TFR is difficult as both the luminosity and the angular momentum might 
be evolving at the same time. For this reason, the stellar mass TFR, 
which correlates the stellar mass and the maximum rotational velocities of disks, 
offers a more physically robust comparison as it involves more fundamental 
quantities (Bell \& de Jong \citealp{belldj}, McGaugh et al. \citealp{mcgaugh}, 
Pizagno et al. \citealp{pizagno}). 
In this context, Flores et al. (\citealp{flores}), Kassin et al (\citealp{kassin})
and Conselice et al. (\citealp{conselice}) have found no evolution 
in both the slope and zero point of the stellar mass TFR up to $z\sim1$. 
On the other hand, Puech et al. (\citealp{puech}) detect an evolution of the stellar 
mass TFR zero point of 0.36 dex between $z\sim0.6$ and $z=0$, using a sample of 18 disk-like 
galaxies observed with the integral field spectrograph GIRAFFE at the VLT. The differences with 
the results obtained by Flores et al. (\citealp{flores}) on a similar dataset are attributed 
to both a difference in the reference local relation and a more accurate measurement of the 
rotation velocity at high redshift.
A very mild evolution was also found in cosmological N-body/hydrodynamical simulations 
of disc galaxy evolution (Portinari \& Sommer-Larsen \citealp{portinari}, see 
Sect.~\ref{obsTF}), where the galaxy growth with time since $z\sim1$ is happening mostly 
along the TFR. The growth of the disk baryonic mass by infall of gas 
and mergers between $z\sim0-1$ is therefore observed to be accompanied by a correspondingly large
increase of the rotation velocity, so that the evolution of individual objects occurs mainly 
along the TFR. 

\subsection{The observed $z\sim2$ Tully-Fisher relation} \label{obsTF}

With the available modeled kinematics of $z\sim 2$ galaxies from our
SINS data, we are now able to push, for the first time with a sizeable sample, 
the investigation of the evolution of the TFR to even higher redshift. 
When comparing the evolution of the high-z TFR with that inferred from local 
(or lower redshift) samples of star forming galaxies, one must be aware that we 
are comparing different classes of objects, which are not necessarily linked 
from an evolutionary point of view. Therefore any evolution in the relation
can not be interpreted straightforwardly as evolution of any individual galaxy,
as from $z = 2$ to $z = 0$ part of our galaxies may
   undergo significant (morphological/dynamical) evolution
   driven by efficient secular/internal dynamical processes
   and/or merging (cf. Genzel et al. \citealp{genzel08}, Genel et al. \citealp{shy}).
   Rather, evolution of TFR represents the evolution
   of ensemble properties of disk-like populations at different
   redshifts. 

\begin{figure}
	\centering
	{\includegraphics[width=10cm,keepaspectratio]{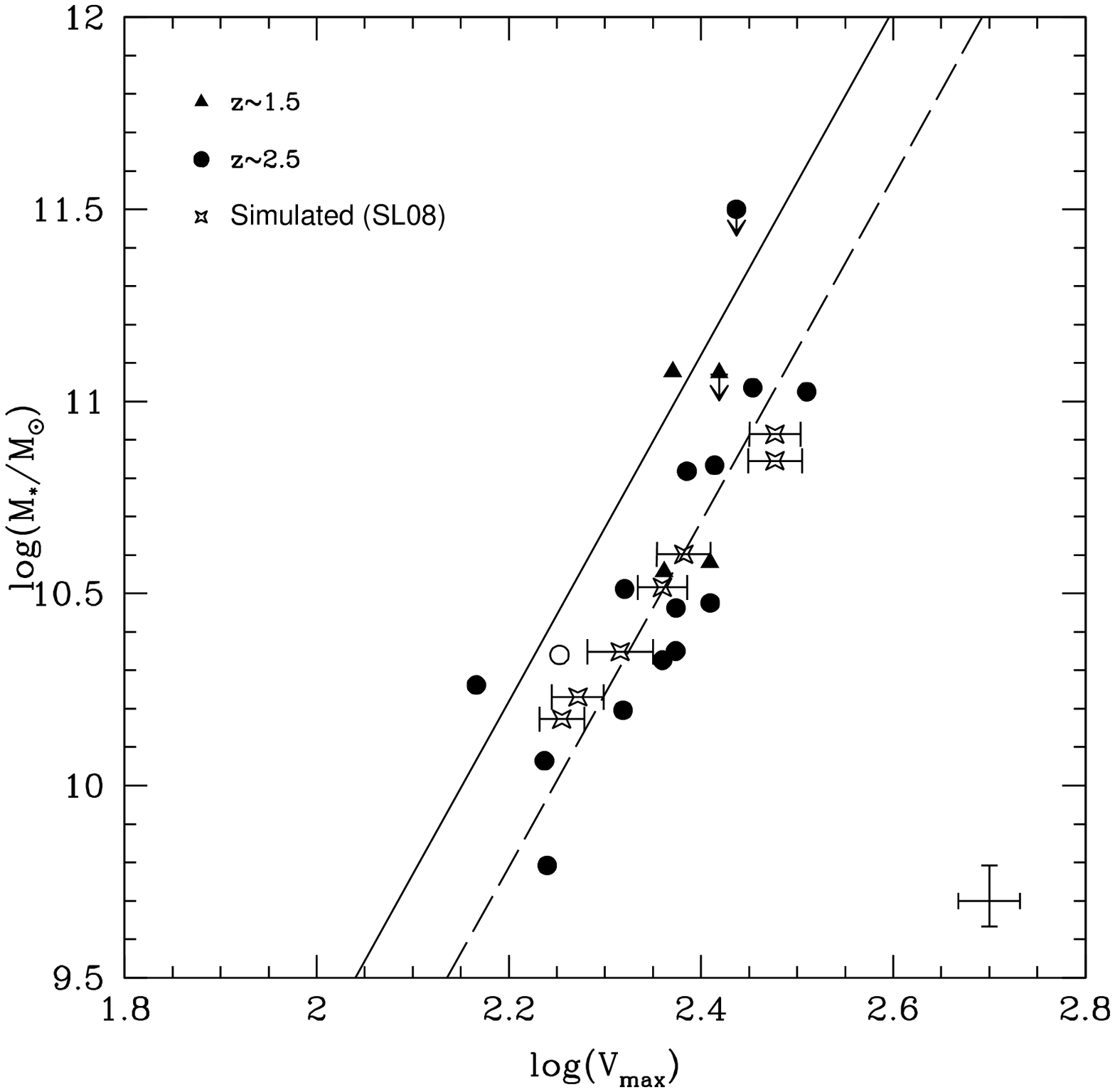}}
	\caption{The stellar mass Tully-Fisher relation at $z\sim 2$. The filled 
	triangles are the $z\sim1.5$ galaxies from our sample, while the filled circles are 
	at $z\sim2.2$. The empty circle show the $z=2.03$ disk galaxy F257 observed by 
	van Starkenburg et al. (\citealp{vanstark}). 
	The error bars in the lower right corner represent the 
	average fitting uncertainties of the model maximum velocity and of the stellar 
	mass from the SED fitting, not accounting for systematic errors.
	The solid line is the $z=0$ relation from Bell \& de Jong (\citealp{belldj}), corrected to a 
	Chabrier IMF. %, and the 
	%dotted lines are the observed scatter at $z=1$ (from Conselice et al. 
	%\citealp{conselice}). 
	The dashed line is instead the best fitting zero point 
	to the $z\sim2.2$ observed galaxies, keeping the slope fixed at the $z\sim0$ value 
	from Bell \& de Jong (\citealp{belldj}). The open diamonds are the $z\sim2.2$ 
	simulated galaxies of Sommer-Larsen et al. (2008), where the bars show the variation 
	in the rotational velocities in the model assuming a $z\sim2.2$ scale length ranging 
	from the $z=0$ value to half of it.
}
	\label{smTF}
\end{figure}

The available full 3D coverage of the dynamical features will not only provide 
a more complete and detailed view of the kinematics, but will also allow us to 
avoid the biases and limitations due to slit spectroscopy, such as 
misalignment of the slit and the major axis, and to identify disturbed rotational 
patterns avoiding the contamination of our disk-like sample with more complex dynamics. 
Using similar 3D SINFONI data, Starkenburg et al. (\citealp{vanstark}) have presented 
the position of a single galaxy at $z=2.03$ on the stellar mass TFR, yielding an offset 
of ($\sim 2\sigma$) from the local one (see Fig.~\ref{smTF}). 
In view of the uncertainties of the quantities involved, which are not insignificant at high-z, 
larger samples are required to constrain more robustly any offset in the TFR at $z\sim2$.

We have therefore used the best fitting V$_{max}$ from our dynamical modeling, together 
with the stellar mass from SED fitting discussed in sect.~\ref{stmass}, to build 
the TFR for our sample, shown in Fig.~\ref{smTF}.
The solid line shows the $z=0$ best fit relation obtained by Bell \& de Jong 
(\citealp{belldj}). Both are corrected to convert their ``diet'' 
Salpeter  and Salpeter IMF to the Chabrier IMF used here, using a factor 1.19 and 1.7 respectively. 
We note that Pizagno et al. (\citealp{pizagno})  have obtained a 
shallower ($M_* \propto V^{3.05}$) local stellar mass TFR than Bell \& de 
Jong (\citealp{belldj}, $M_* \propto V^{4.5}$), but the two relations agree at $V=200\
km\ s^{-1}$. However, they used the circular velocity at $2.2\ R_d$ instead of $V_{max}$ to 
estimate of the rotational velocity, and differences in the sample selection may also 
account for the different result.
McGaugh et al. (\citealp{mcgaugh}) obtained an intermediate slope, $M_{bar} \propto V^{4.0}$, but 
they explored the total baryonic TFR, including the contribution of the gaseous mass 
component. In this case, higher gas fraction in lower mass galaxies may explain the flatter slope.
Recently Meyer et al. (\citealp{meyer}) presented the local stellar mass TFR for a large 
HI-selected sample of galaxies, obtaining a slope consistent to the Bell \& de Jong (\citealp{belldj})
value adopted here for comparison with previous studies.

The circles in Fig.~\ref{smTF} show the data for the $z\sim2.2$ galaxies from SINS, while the 
filled triangles the $z\sim1.5$ sources. The open circle is the $z\sim2$ galaxy presented in 
van Starkenburg et al. (\citealp{vanstark}).
%At $z\lesssim 1$ the TF 
%scatter was found to be partly due to disturbed and peculiar morphologies, corresponding to lower 
%velocities for their masses (see Kassin et al. \citealp{kassin}).
The average uncertainty is shown as an error bar in the lower 
right corner. This does not include possible systematic effects both in the SED fitting 
(see Sect.~\ref{stmass}) and in the dynamical modeling (see Sect.~\ref{modeling}).
The $z\sim2.2$ galaxies show an evident evolution in the zero point of the TFR. 
The dashed line shows the best fitting relation 
keeping the slope fixed at the $z=0$ value found by Bell \& de Jong (\citealp{belldj}). 
The current data are in fact not yet able to fully constrain the slope 
of the relation due to the limited statistics and mass range. 
The best fitting zero point is $ZP_{2.2}=-0.09\pm0.11$, where the error is the $1 \sigma$ uncertainty, 
to be compared with the local $ZP_{0}=0.32$, giving a significance of $\sim 3.7\sigma$.
The $z=2.2$ relation can be written as $log(M_*)=-0.09 +4.5 \times \log(V_{max})$, finding 
%that $M_{*} \sim 0.30$ times the one  at $z\sim0$ for a given rotational velocity.
an offset in $log(M_*)$ of $0.41$ for a given rotational velocity.
We do not find evidence for significant evolution in our $z\sim1.5$ sample with respect to the 
local relation, although the data are still too sparse in this redshift range to place 
a firm constraint. 
The evolution observed at $z\sim2.2$ is larger than the intermediate 0.36 dex 
found at $z\sim0.6$ by Puech et al. (\citealp{puech}), although they use a different local reference, 
suggesting an increasing of the evolution of the TFR with redshift.

\subsection{Comparison with galaxy evolution models}

Interestingly, the amount of observed evolution in our sample is consistent with the factor $\sim1.25$ 
of increased velocity for a given stellar mass predicted by Somerville et al. 
(\citealp{somerville}) in their semi-analytic formation model 
of a disk with $M_*=6\times 10^{10}\ M_{\odot}$ at this redshift. 
They follow the classical paradigm for disk 
formation within massive extended dark matter halos initially proposed by Fall 
\& Efstathiou (\citealp{fall}) but use a Navarro et al. (\citealp{nfw}) 
mass density profile instead of a singular isothermal sphere as halo mass density profile, 
account for the expected evolution with time of the profile concentration $c_{vir}$ 
(Bullock et al. \citealp{bullock}), and for the effect of self-gravity of the baryons in the central 
part of the halo (so called ``adiabatic contraction'').  They find that with this refined model 
the disk rotation velocity $V_{max}$ at fixed disk mass evolves much more gradually than the halo 
virial velocity at fixed halo mass, remaining nearly constant and showing the largest evolution 
between $z\sim1.5-3$. If this prediction of the model 
is correct, our sample therefore probes the best suited cosmic epoch to test the evolution of 
the TFR.

We also compare the observed Tully-Fisher relation with the Smoothed Particle Hydrodynamic (SPH) 
simulations of disk formation and 
evolution in a cold dark matter hierarchical clustering scenario by Sommer-Larsen et al. 
(in preparation) and Portinari \& Sommer-Larsen (\citealp{portinari}). They include in their 
modeling star formation and stellar feedback effects to delay the collapse of protogalactic gas 
clouds and allow the infalling gas to preserve a large fraction of its angular momentum. Previous 
numerical simulations have in fact shown that when only cooling processes are included, the infalling 
gas loses too much angular momentum, resulting in disks that are much smaller than the observed 
ones (Navarro \& White \citealp{navarroW}, Navarro \& Steinmetz \citealp{navarroS}).
Moreover, chemical evolution with non-instantaneous recycling, metallicity-dependent radiative 
cooling and the effects of a meta-galactic ultraviolet field, including simplified radiative transfer,
were implemented in the simulations. This is an improved version of the models developed by 
Sommer-Larsen et al. (\citealp{sl99}, \citealp{sl03}) to address the angular momentum problem 
in disk formation models and provide realistic $z=0$ galaxies. It has proven successful in 
reproducing better than previous numerical models the observed sizes of disk galaxies, the 
luminosity profile of both disks and bulges, bulge to disk ratios, integrated colors and TFR 
of local galaxies, as well as the observed peak in the cosmic star formation rate at $z\sim2$ 
(Sommer -Larsen et al. \citealp{sl03}).
The models are characterized by a formation mechanism  where a large 
fraction of the gas is accreted onto the forming disk through both rapid cold gas accretion 
in filamentary structures (dominant at $z\gtrsim2$) and gradual cooling flows 
from a surrounding hot phase (dominant for $z\lesssim2$), without abrupt changes of angular momentum 
orientation, rather than by rapid mergers of massive cold clumps. 
These simulated galaxies at $z\sim2.2$ are shown as open diamonds in Fig.~\ref{smTF}. Again, we 
obtain very good agreement between our data and the predicted position of $z\sim2.2$ galaxies 
on the TFR for a range of stellar masses. In the model the predicted zero point offset 
is produced by a relevant inside-out mass accretion between $z\sim2.2$ and $z\sim1$, 
that do not correspond to an equal increase of the rotation velocity of the galaxies. This issue 
will be investigated in a forthcoming paper (Sommer-Larsen et al. in preparation).
%We stress here that these simulations are able to reproduce both the local TFR and the very small 
%evolution observed at $z\sim1$, as well as the evolution detected in our higher redshift data,  
%providing strong support in favor of a scenario in which rapid cold gas accretion 
%in filamentary structures from the halo is required to reproduce the observed properties of the 
%$z\sim2$ galaxies.
However, if the TFR evolution is driven mainly by dark matter as suggested by the models, 
this may suggest that the galaxies we are observing at high-z are ``stable'', in contradiction with inferences from 
our detailed studies of several of our disks which appear to be globally unstable to star formation 
(see Genzel et al. \citealp{genzel08}).

\subsection{The $S_{0.5}$ Tully-Fisher relation}

\begin{figure}
	\centering
	{\includegraphics[width=10cm,keepaspectratio]{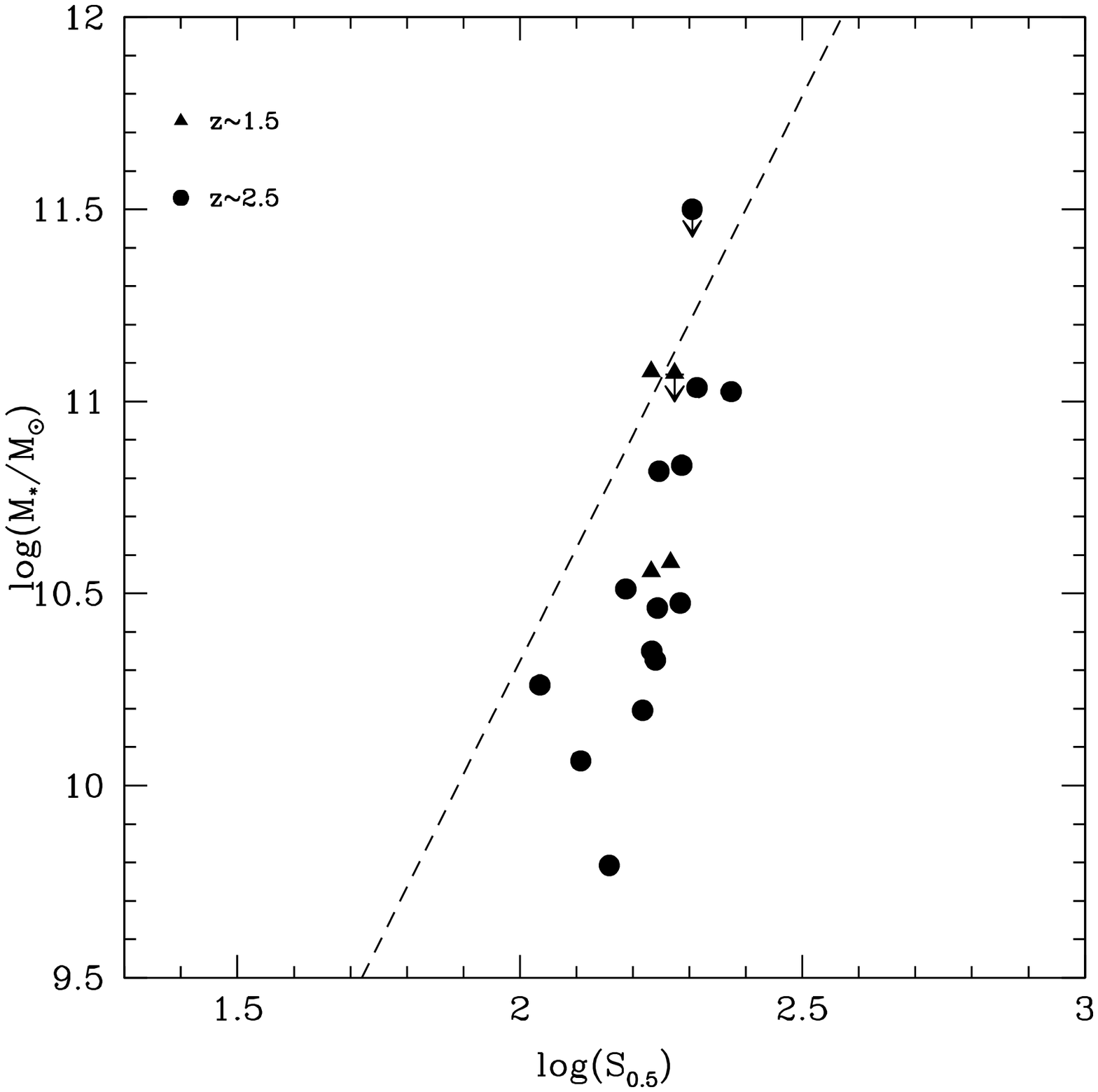}}
	\caption{The stellar mass Tully-Fisher relation at $z\sim 2$ using the 
	$S_{0.5}=\sqrt(0.5\cdot V^2_{rot}+\sigma_0^2)$ estimator introduced by 
	Kassin et al. (\citealp{kassin}), in order to take into account disordered 
	or non-circular motions through the gas. The filled  
	triangles are the $z\sim1.5$ galaxies from our sample, while the circles are 
	at $z\sim2.2$. The dashed line is the $z\sim1$ relation derived by 
	Kassin et al. (\citealp{kassin}).
	%The error bars in the lower right corner represent the 
	%average fitting uncertainties of the model maximum velocity and of the stellar 
	%mass from the SED fitting, not accounting for systematic errors.
	}
	\label{S05TF}
\end{figure}

Kassin et al. (\citealp{kassin}) have constructed a stellar mass TFR 
at $z\lesssim1$ using the $S_{0.5}=\sqrt{0.5\cdot V^2_{rot}+\sigma_{0}^2}$ estimator, 
in order to take into account disordered or non-circular motions through the gas.
In this way, they found a tighter relation up to $z=1.2$, with no significant 
evolution, and conclude that the observed scatter in the TFR is partly due to 
disturbed and peculiar morphologies, corresponding to lower rotation  
velocities for their masses. In Fig.~\ref{S05TF} the result obtained for our $z\sim2$ sample using 
the $S_{0.5}$ index is shown, using the fitted $\sigma_{02}$ as a measure of the 
intrinsic dispersion of the gas not due to rotation. Although the $\sigma$ used here does not 
exactly match the definition adopted by Kassin et al. (\citealp{kassin}), it is the best 
indicator of turbulent and disturbed motion obtained from our modeling. Using the $S_{0.5}$ 
index we also detect an evolution of the relation between $z\sim1$ and $z\sim2.2$, further supporting 
the claim of an evolution of the TFR. However, due to the different definition of $\sigma$ used here, 
we do not attempt to quantify this evolution. The resulting scatter is slightly smaller than the 
already remarkably low one obtained for the classical TFR (see Fig.~\ref{smTF}). We suggest that 
this is due to the optimal selection of our rotational supported systems at $z\sim2$, as the 
SINFONI 3D data are ideal to ensure a much lower contamination from complex dynamics and irregular 
motions than in longslit datasets. 

\section{Conclusions} \label{conclusions}

We have presented the dynamical modeling of the full 2D velocity and velocity dispersion 
fields of the H$\alpha$ emission line for 18 star forming galaxies at $z\sim2$, observed
with the integral field spectrometer SINFONI in the near-IR. The galaxies are selected from 
the larger sample of the SINS survey, which probes galaxy populations responsible for the
bulk of the cosmic star formation activity and stellar mass density at these redshifts, 
based on the prominence of ordered rotational motions with respect to more complex merger 
induced dynamics. The quality of the data allows us to model the observed dynamics with 
suitable simulated exponential disks, and derive the total dynamical mass and maximum 
rotational velocity of these galaxies. 

We find a good correlation between the dynamical mass and the stellar mass derived 
through SED fitting, deriving that large gas fractions ($M_{gas}\approx M_*$) are required to 
explain the difference between the two quantities. 
We also find that the galaxies with younger stellar population age have generally larger 
gas fractions, as reported by Erb et al. (\citealp{erbb}), 
although the large observed spread could be related to a significant 
contribution of continuous gas accretion.

We used the derived stellar mass and the $V_{max}$ of the best fitting disk model of 
each galaxy to construct for the first time a Tully-Fisher relation at $z\sim2.2$. This is 
the redshift range where the largest evolution may be expected according to some recent
simulations (Somerville et al. \citealp{somerville}). The 
full 2D spatial coverage of our data provides a view of the kinematics in unprecedented detail at 
these redshifts, with which we can avoid the biases and limitations of slit spectroscopy 
in studying the dynamics of high-z sources. Moreover, the full 2D dynamical information 
allows us to select a sample with a lower contamination from irregular and complex dynamics, 
which is better suited for studying the evolution of the TFR in rotationally dominated systems. 
We find a remarkably low scatter in the Tully-Fisher relation
in our sample, and an evolution of the zero point in $log(M_*)$ of $0.41\pm0.11$ 
for a given rotational velocity, which is 
fully consistent with the predictions of the latest N-body/hydrodynamical simulations of disks 
formation and evolution. To reproduce the observed properties of the local and high-z TFR, 
the SPH model used as comparison (Sommer-Larsen et al. \citealp{sl99}, 
\citealp{sl03} and in preparation) invokes a formation mechanism where a large 
fraction of the gas is accreted onto the forming disk through both rapid cold gas accretion 
in filamentary structures and gradually in a cooling flow 
from a surrounding hot phase, rather than by rapid mergers of massive cold clumps. 
This is in qualitative agreement with recent 
results showing that smooth gas accretion from the halo is required to reproduce the observed 
properties of a large fraction  
of the $z\sim2$ galaxies (see e.g. F\"orster Schreiber et al. \citealp{natascha};
Genzel et al. \citealp{genzel}, \citealp{genzel08}; Shapiro et al. \citealp{kristen};
Noeske et al. \citealp{noeske}; Elbaz et al. \citealp{elbaz}; Daddi et al. \citealp{daddi07b}). 
As already suggested by the velocity-size relation obtained for a larger sample of SINS 
UV/optically selected galaxies (Bouch\'e et al. \citealp{nicolas}), the angular momentum 
properties of these $z\sim2$ galaxies are very similar to local spirals and 
intermediate redshift disks. However, these $z\sim2$ disks show a much larger 
component of random local gas motion ($V/\sigma_{02}\sim4.4$, and up to an order of magnitude 
less than $z=0$ spirals), implying high turbulence and scale heights. The random 
motions could be the result of feedback from the intense star formation ($\sim 50-200\ 
M_{\odot}/yr$), or be driven by the accreting gas as it settles onto the forming disk (see Genzel 
et al. \citealp{genzel08}).

The results obtained confirm that the capabilities of Integral Field Spectroscopy can 
provide unique insights in the formation and evolution of high-z galaxies, and that large, 
gas rich, rotating disks, represent a significant population in the $z\sim2$ universe.

%______________________________________________________________

\acknowledgments

We wish to thank the ESO staff, and in particular at Paranal Observatory
and the SINFONI and PARSEC teams, for their helpful and enthusiastic 
support during the several observing runs over which the SINS program was 
carried out. We would also like to thank the referee for the useful comments 
and suggestions. N. M. F. S. acknowledges support by the Schwerpunkt 
Programm SPP1177 of the Deutsche Forschungsgemeinschaft and by the Minerva 
Program of the Max-Planck-Gesellschaft. 




\clearpage

%% the file, uncomment the \plotone, \plottwo, and
%% If you need a layout that cannot be achieved with \plotone or
%% \plottwo, you can invoke the graphicx package directly with the
%% \includegraphics command or use \plotfiddle. For more information,
%% please see the tutorial on "Using Electronic Art with AASTeX" in the
%% documentation section at the AASTeX Web site,
%% http://www.journals.uchicago.edu/AAS/AASTeX.
%%
%% The examples below also include sample markup for submission of
%% supplemental electronic materials. As always, be sure to check
%% the instructions to authors for the journal you are submitting to
%% for specific submissions guidelines as they vary from

%% This example uses \plotone to include an EPS file scaled to
%% 80% of its natural size with \epsscale. Its caption
%% has been written to indicate that additional figure parts will be
%% available in the electronic journal.

%

\begin{figure}
	\centering
	{\includegraphics[width=12cm,keepaspectratio]{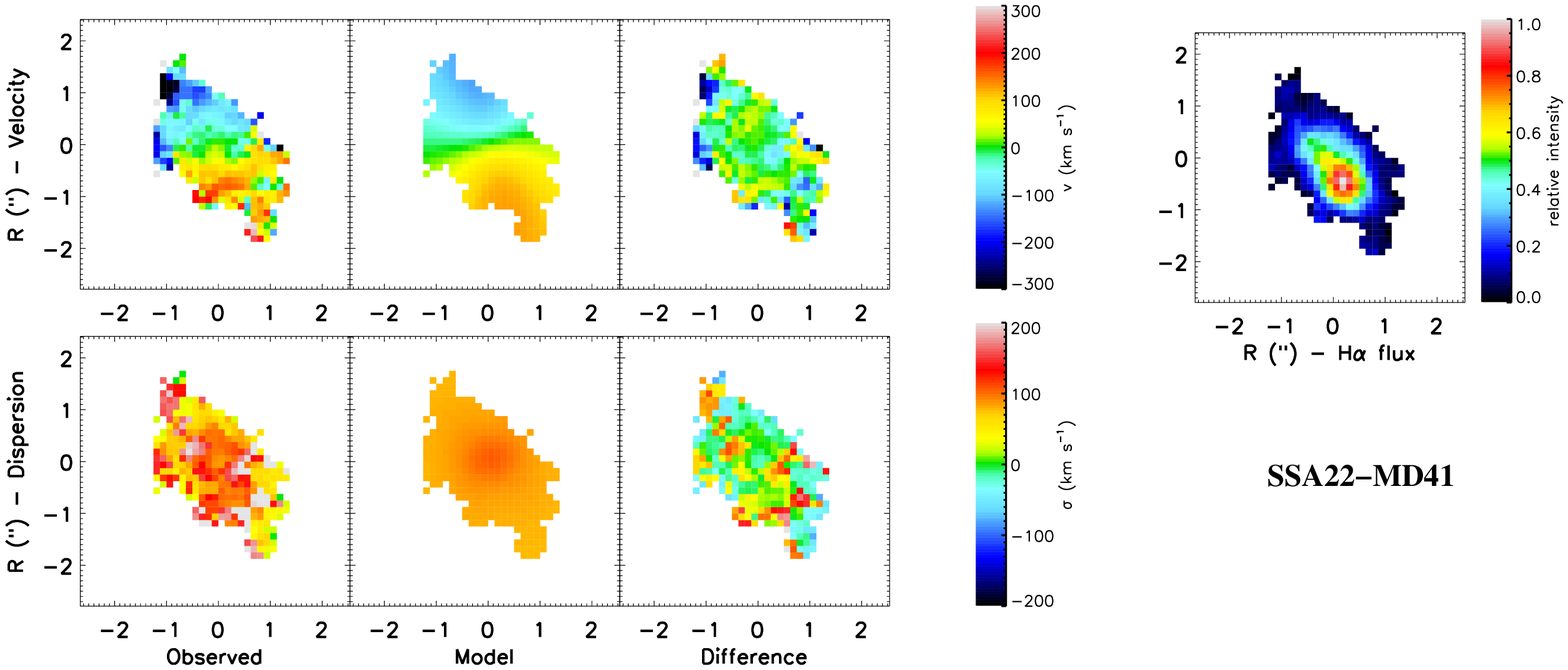}}\\
	\vspace{0.5cm}
	{\includegraphics[width=12cm,keepaspectratio]{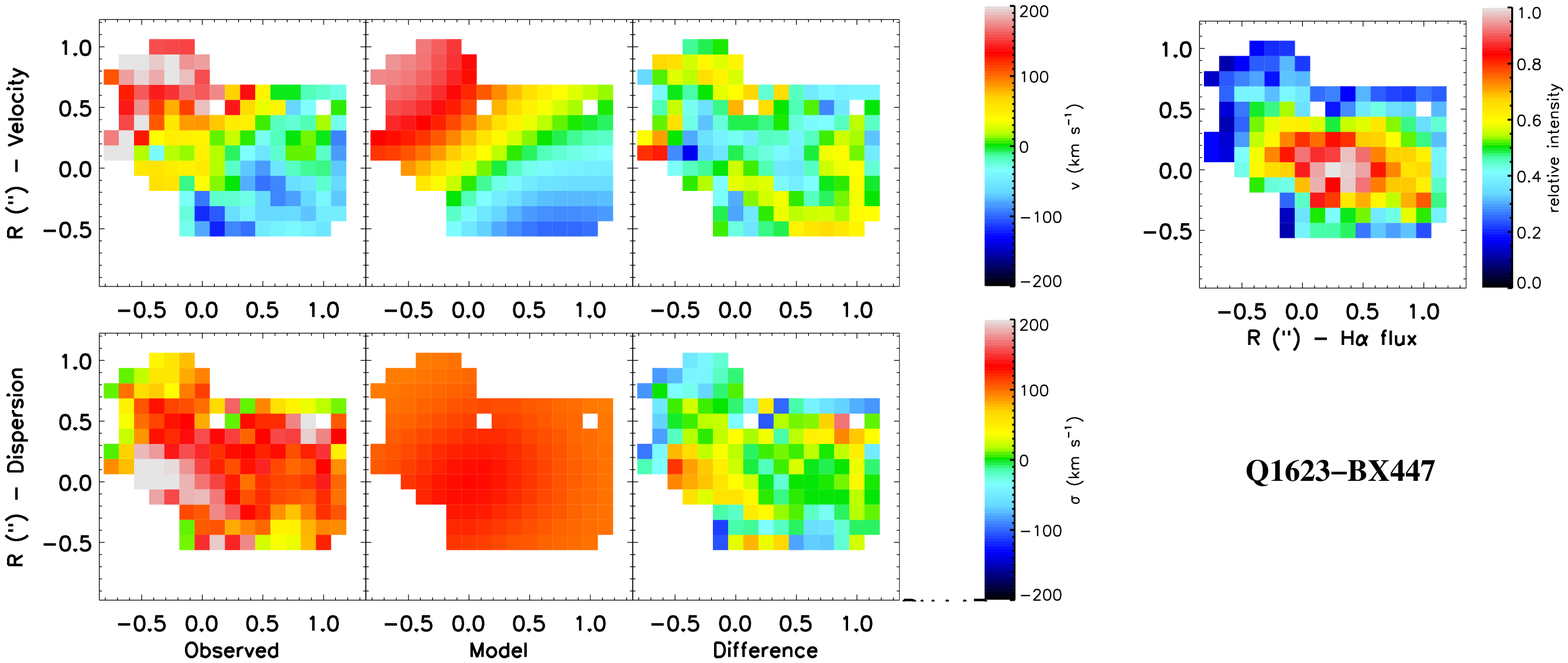}}\\
	\vspace{0.5cm}
	{\includegraphics[width=12cm,keepaspectratio]{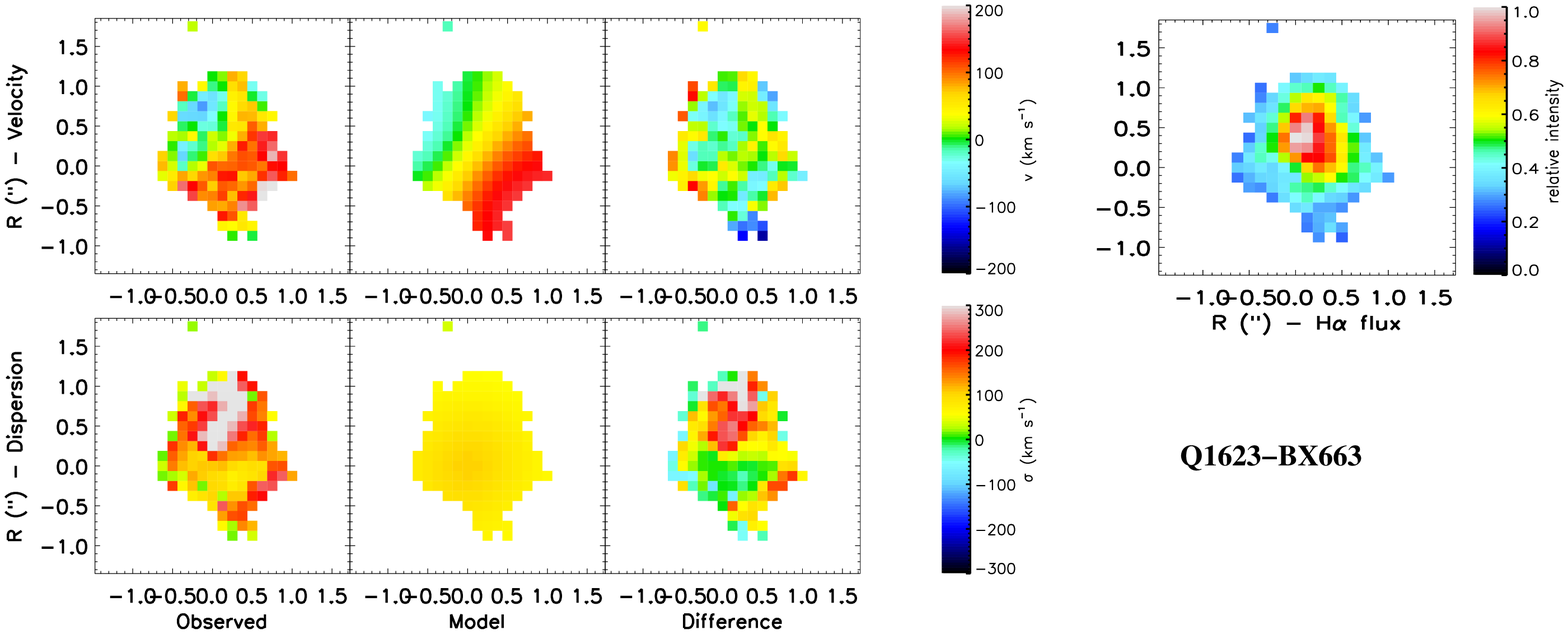}}
	
	\caption{Kinematic fitting of the galaxies of the sample. For both 
	the velocity (upper panel) and velocity dispersion (lower panel), we show from left 
	to right the observed maps, the best fitting model maps and their 
	differences. It can be seen how the global 
	rotational pattern is typically well reproduced, although local deviations from 
	the ideal model are highlighted in the difference. The H$\alpha$ line map is shown in
	the right panel for each galaxy, with the intensity scale normalized to the brightest pixel.}
	\label{exfit}
\end{figure}
	
\begin{figure}
	\figurenum{2}
	\centering
	{\includegraphics[width=12cm,keepaspectratio]{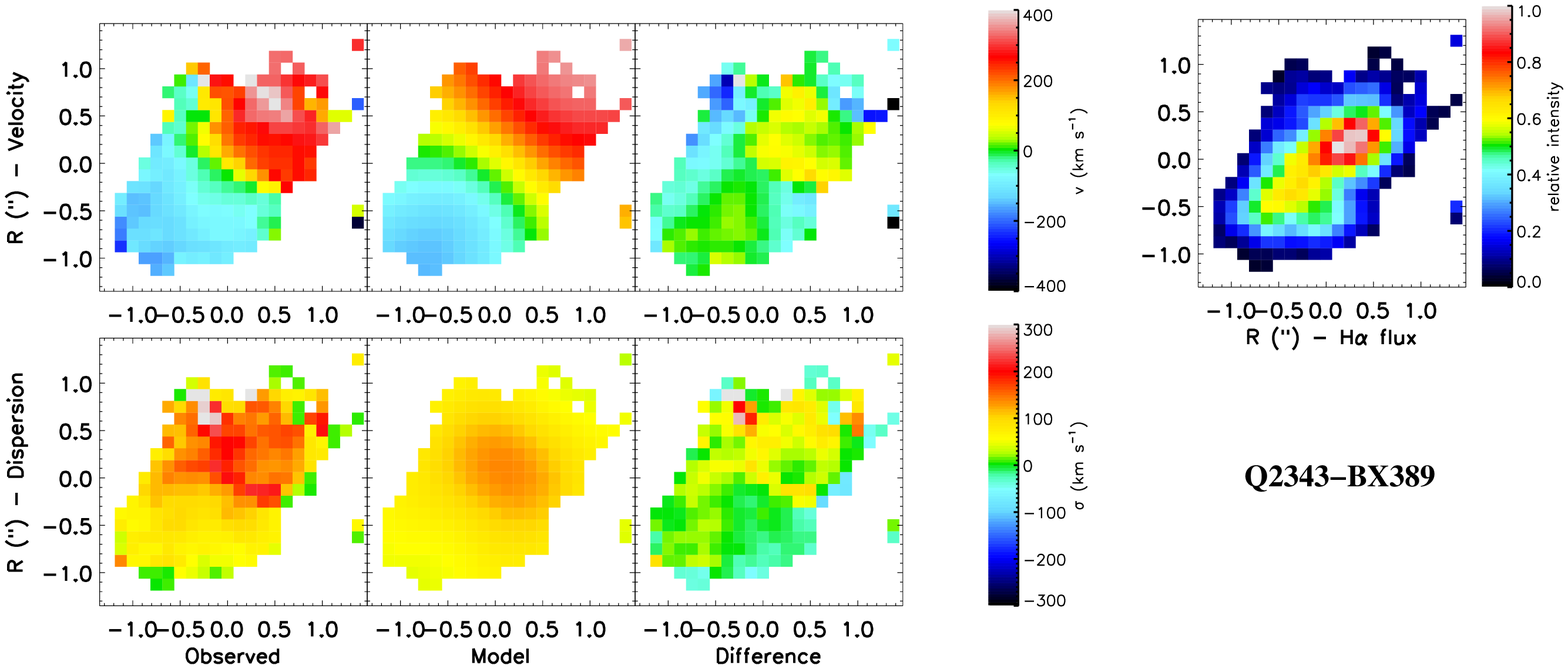}}\\
	\vspace{0.5cm}
	{\includegraphics[width=12cm,keepaspectratio]{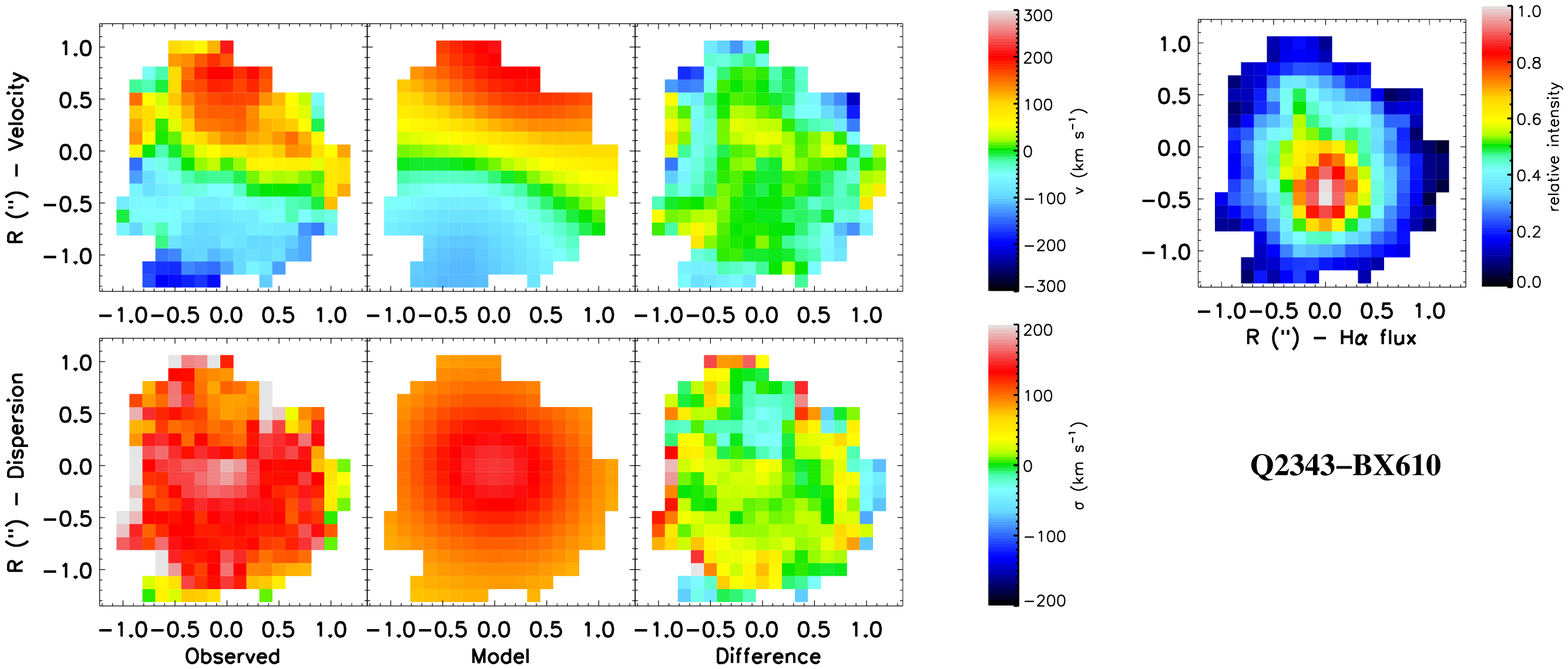}}\\
	\vspace{0.5cm}
	{\includegraphics[width=12cm,keepaspectratio]{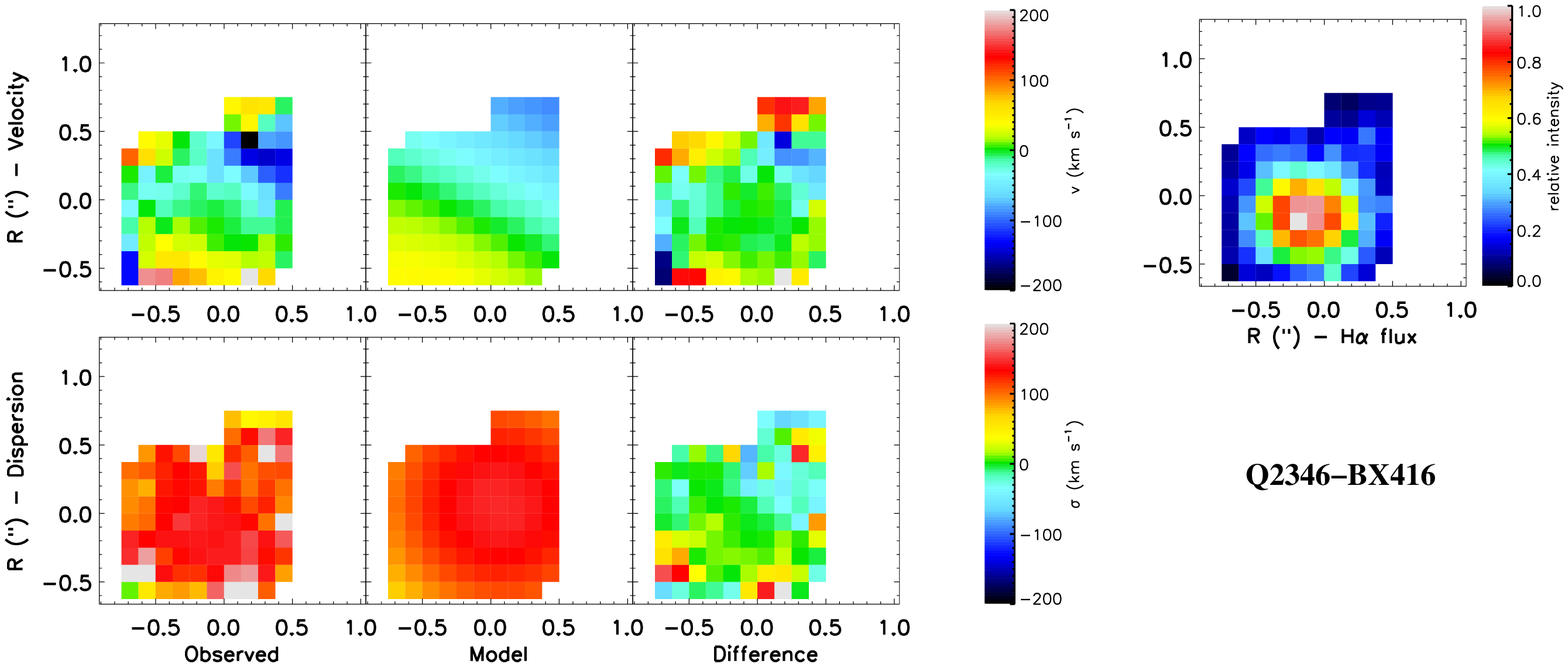}}\\
	\vspace{0.5cm}
	{\includegraphics[width=12cm,keepaspectratio]{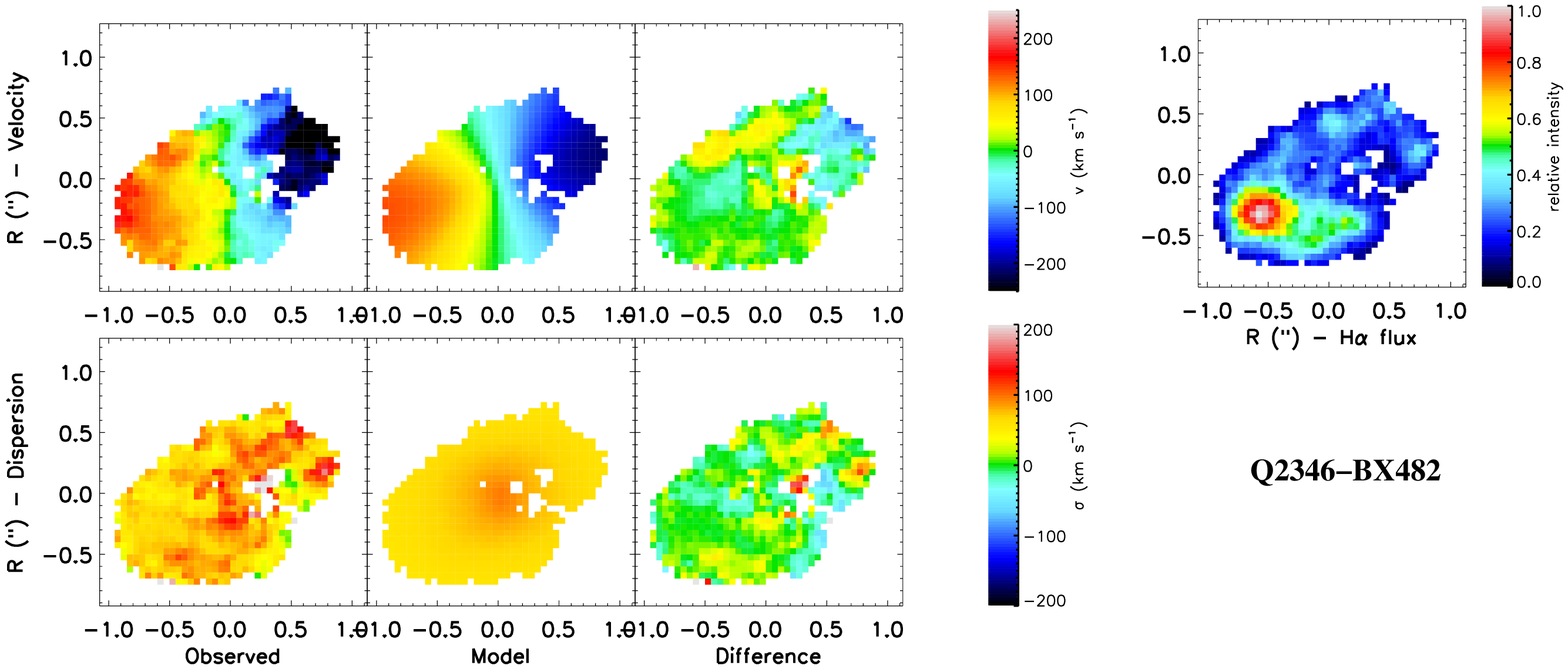}}
	
	\caption{(continued) -- Kinematic fitting of the galaxies of the sample.}
\end{figure}

\begin{figure}
	\figurenum{2}
	\centering
	{\includegraphics[width=12cm,keepaspectratio]{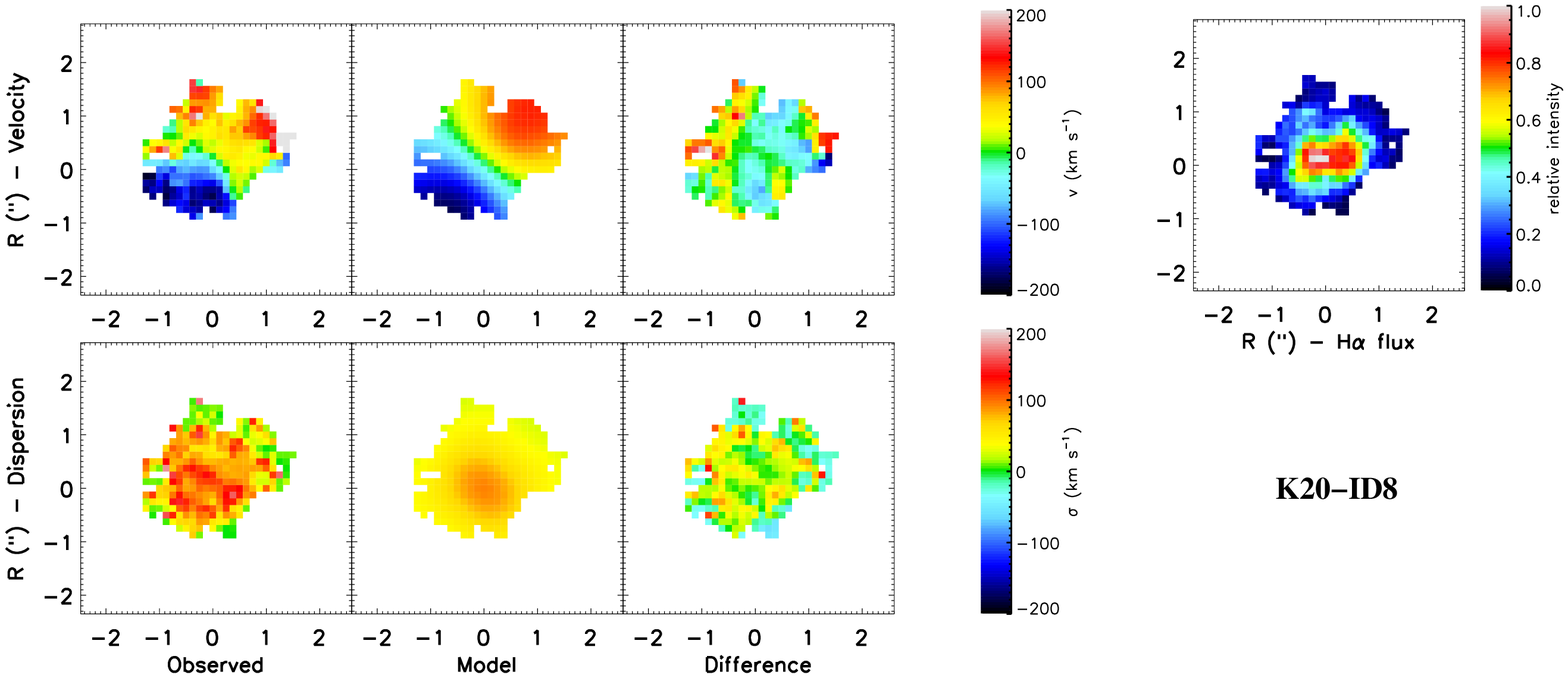}}\\
	\vspace{0.5cm}
	{\includegraphics[width=12cm,keepaspectratio]{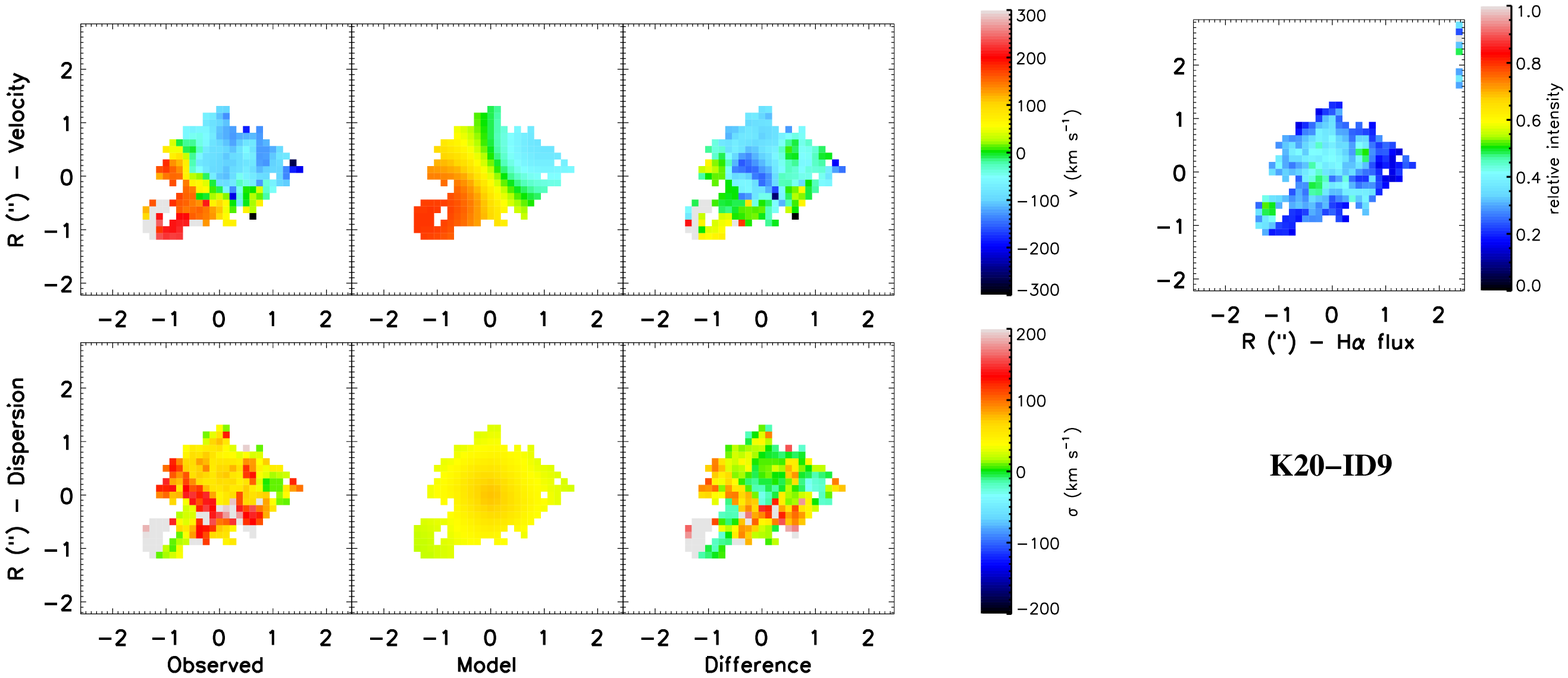}}\\
	\vspace{0.5cm}
	{\includegraphics[width=12cm,keepaspectratio]{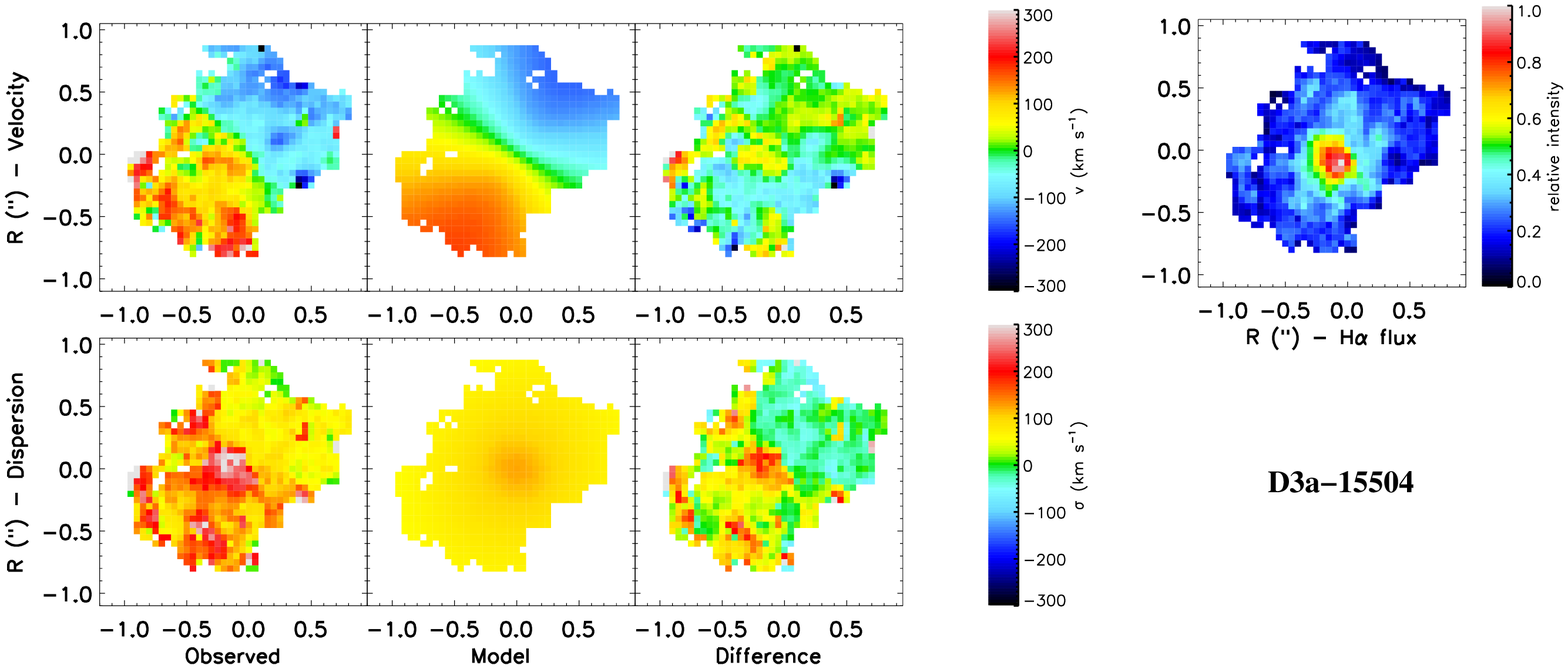}}\\
	\vspace{0.5cm}
	{\includegraphics[width=12cm,keepaspectratio]{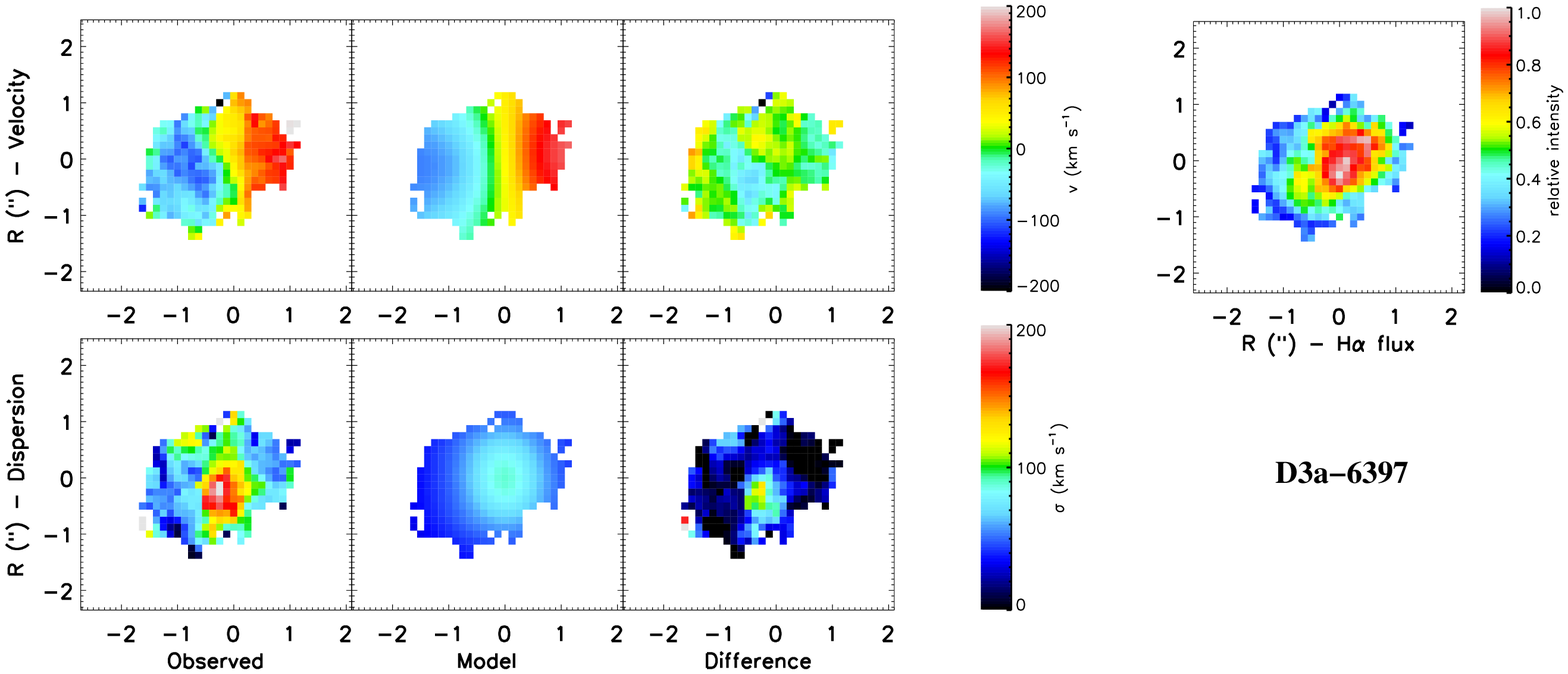}}
	
	\caption{(continued) -- Kinematic fitting of the galaxies of the sample.}
\end{figure}

\begin{figure}
	\figurenum{2}
	\centering
	{\includegraphics[width=12cm,keepaspectratio]{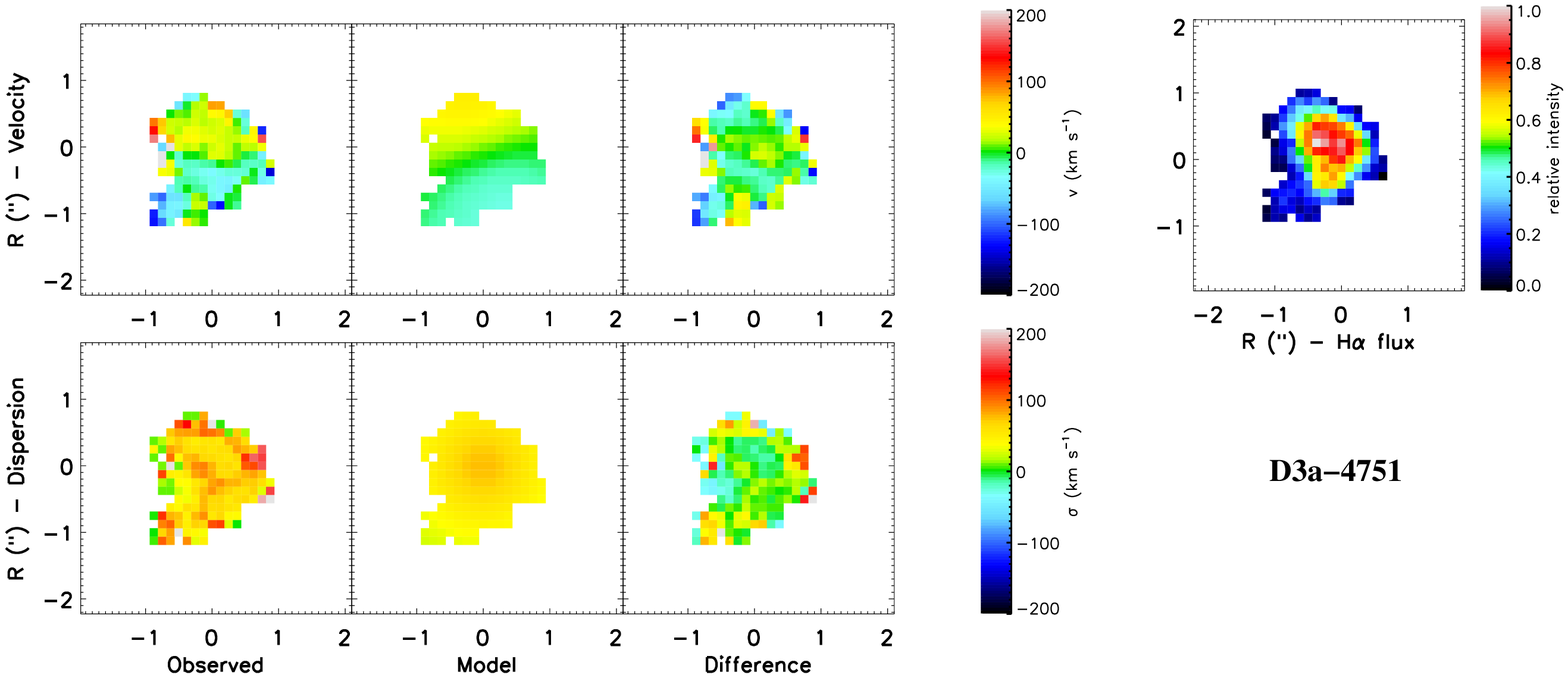}}\\
	\vspace{0.5cm}
	{\includegraphics[width=12cm,keepaspectratio]{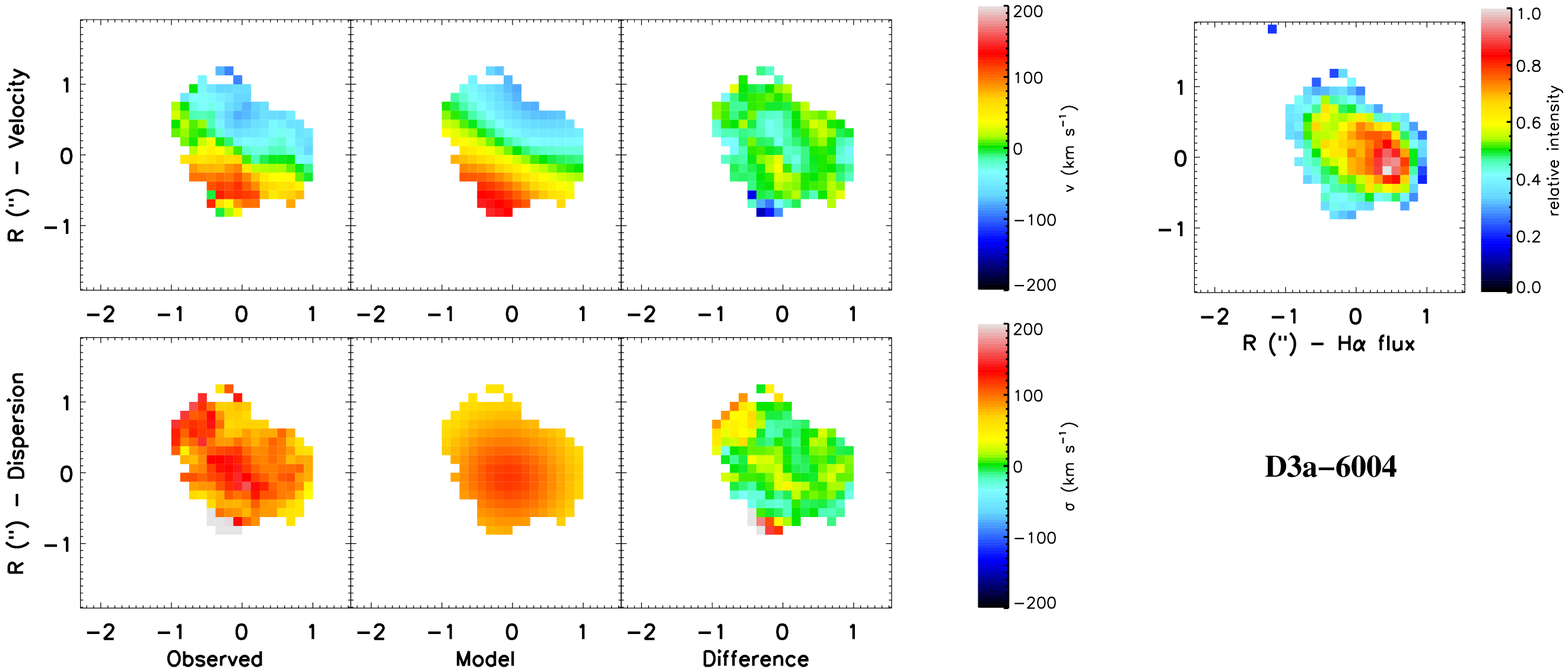}}\\
	\vspace{0.5cm}
	{\includegraphics[width=12cm,keepaspectratio]{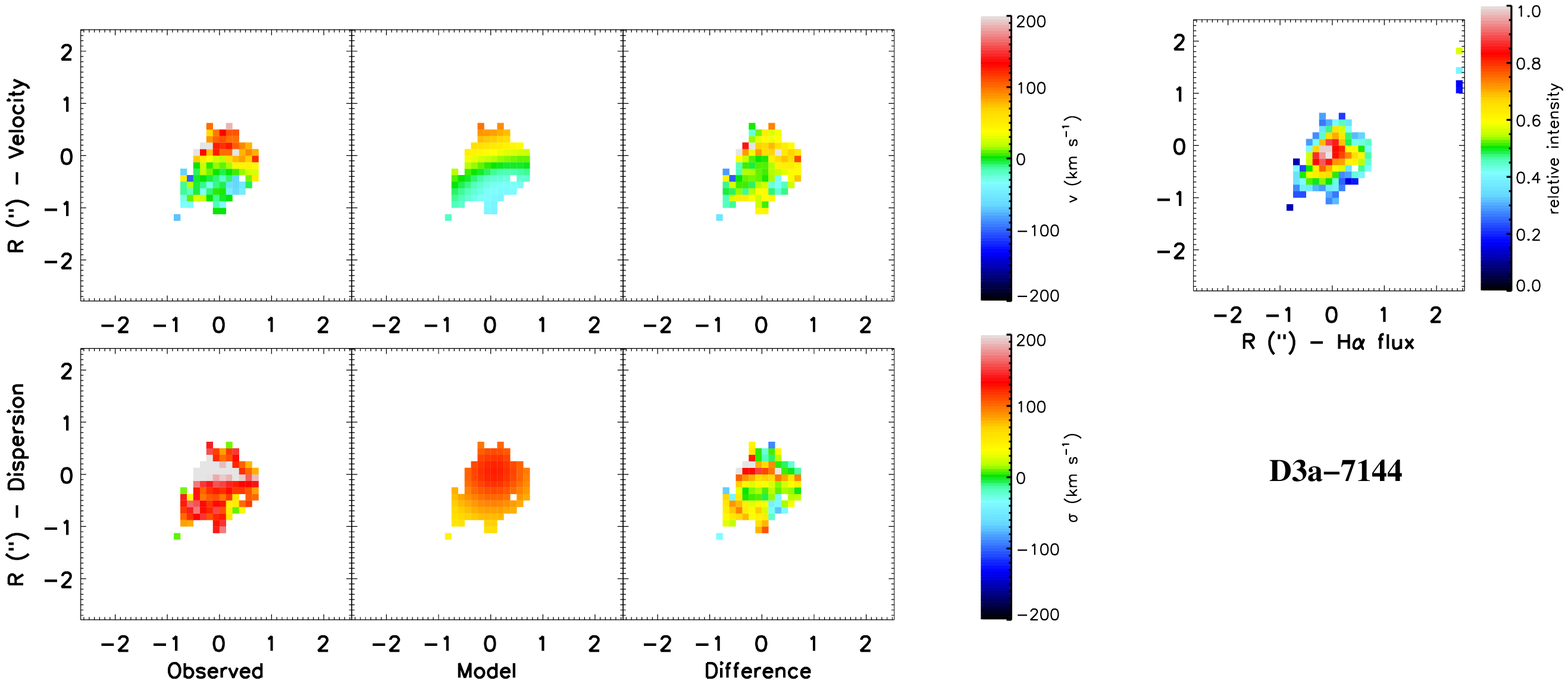}}\\
	\vspace{0.5cm}
	{\includegraphics[width=12cm,keepaspectratio]{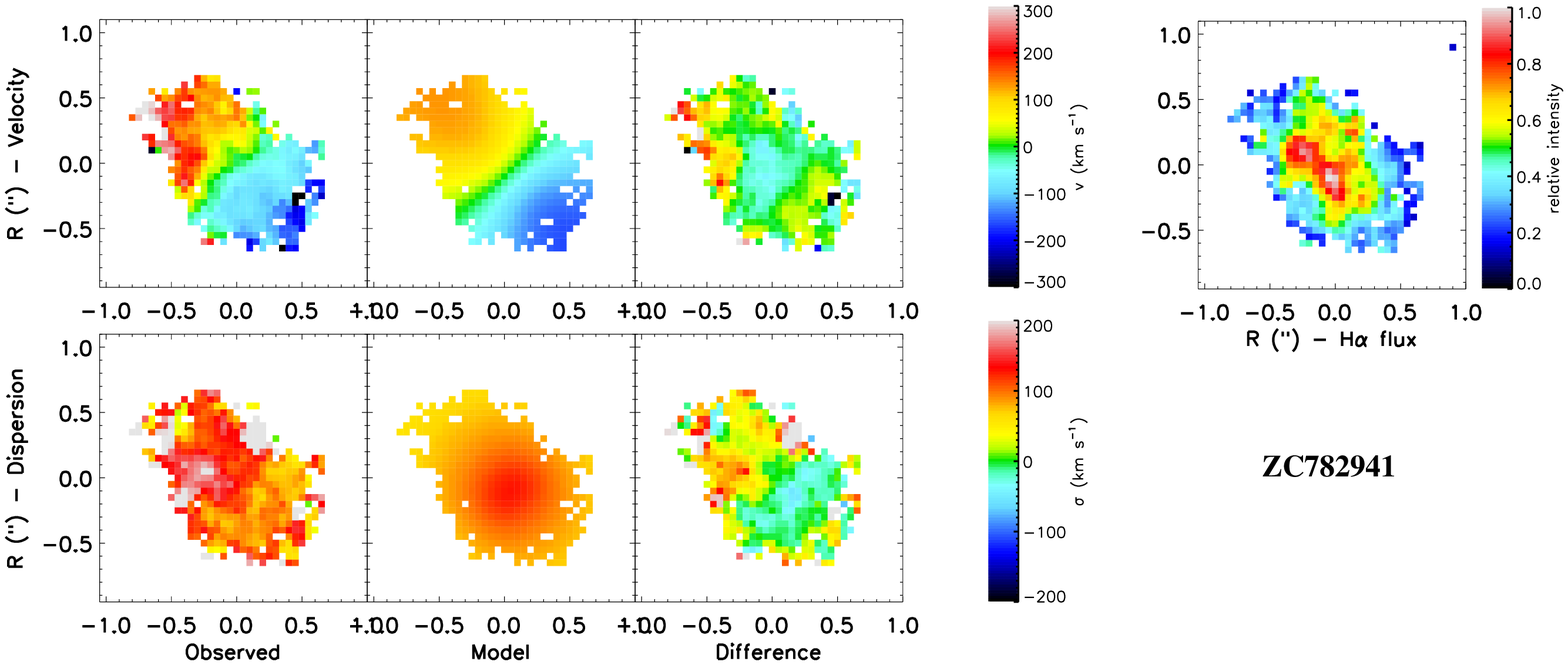}}

	\caption{(continued) -- Kinematic fitting of the galaxies of the sample.}
\end{figure}

\begin{figure}
	\figurenum{2}
	\centering
	{\includegraphics[width=12cm,keepaspectratio]{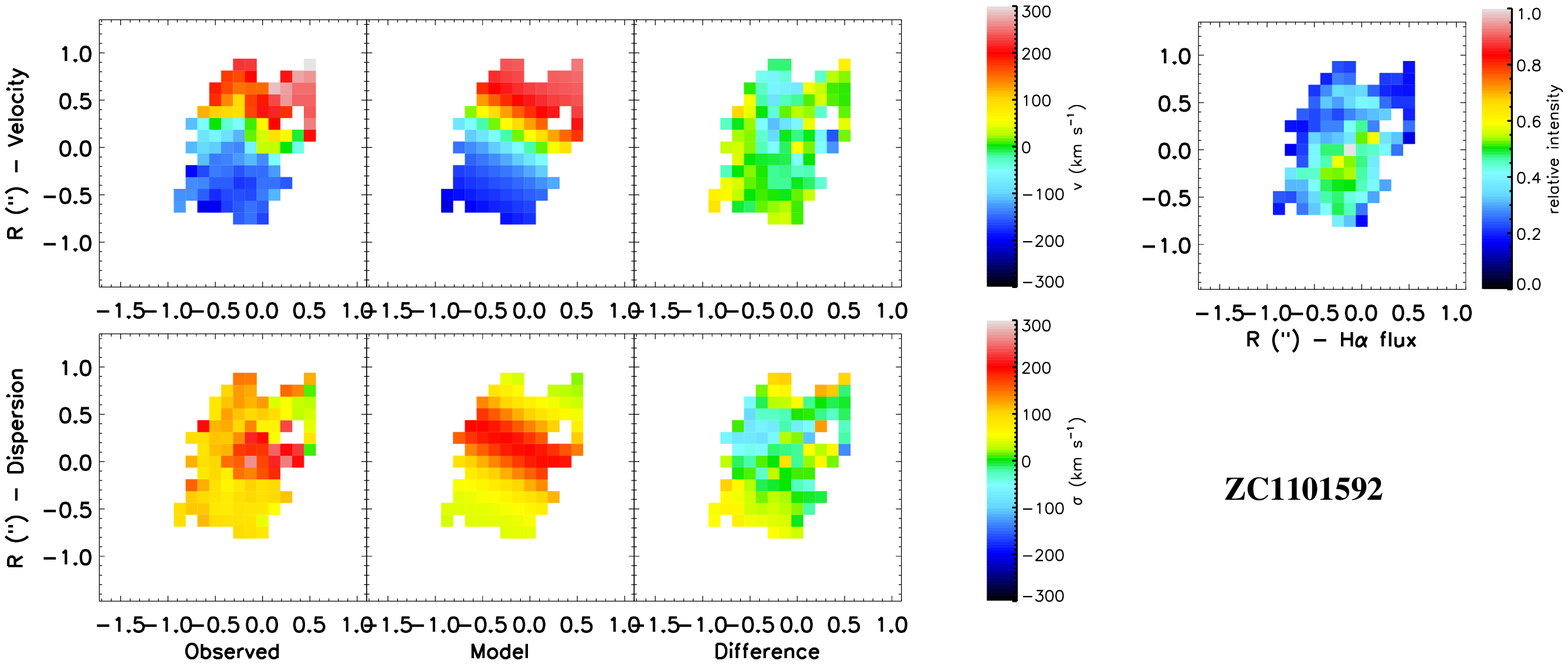}}\\
	\vspace{0.5cm}
	{\includegraphics[width=12cm,keepaspectratio]{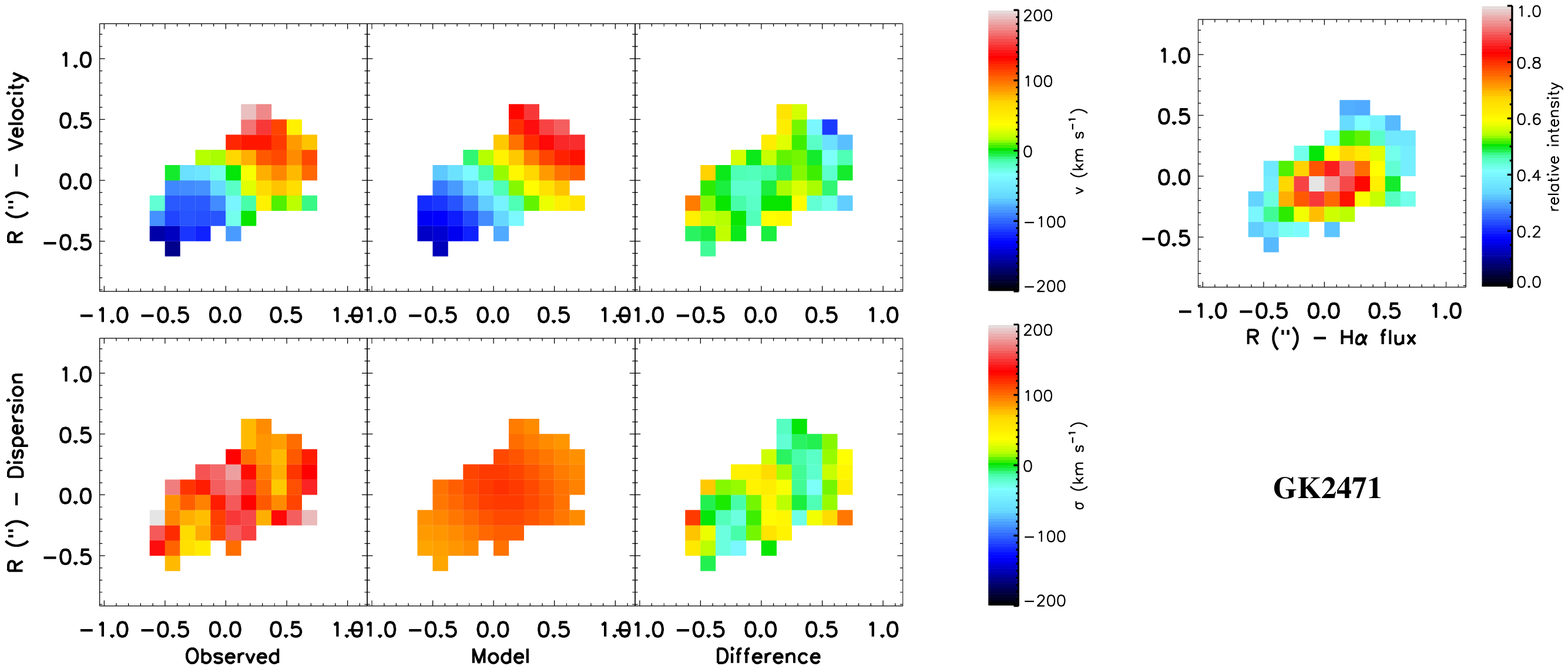}} \\
	\vspace{0.5cm}
	{\includegraphics[width=12cm,keepaspectratio]{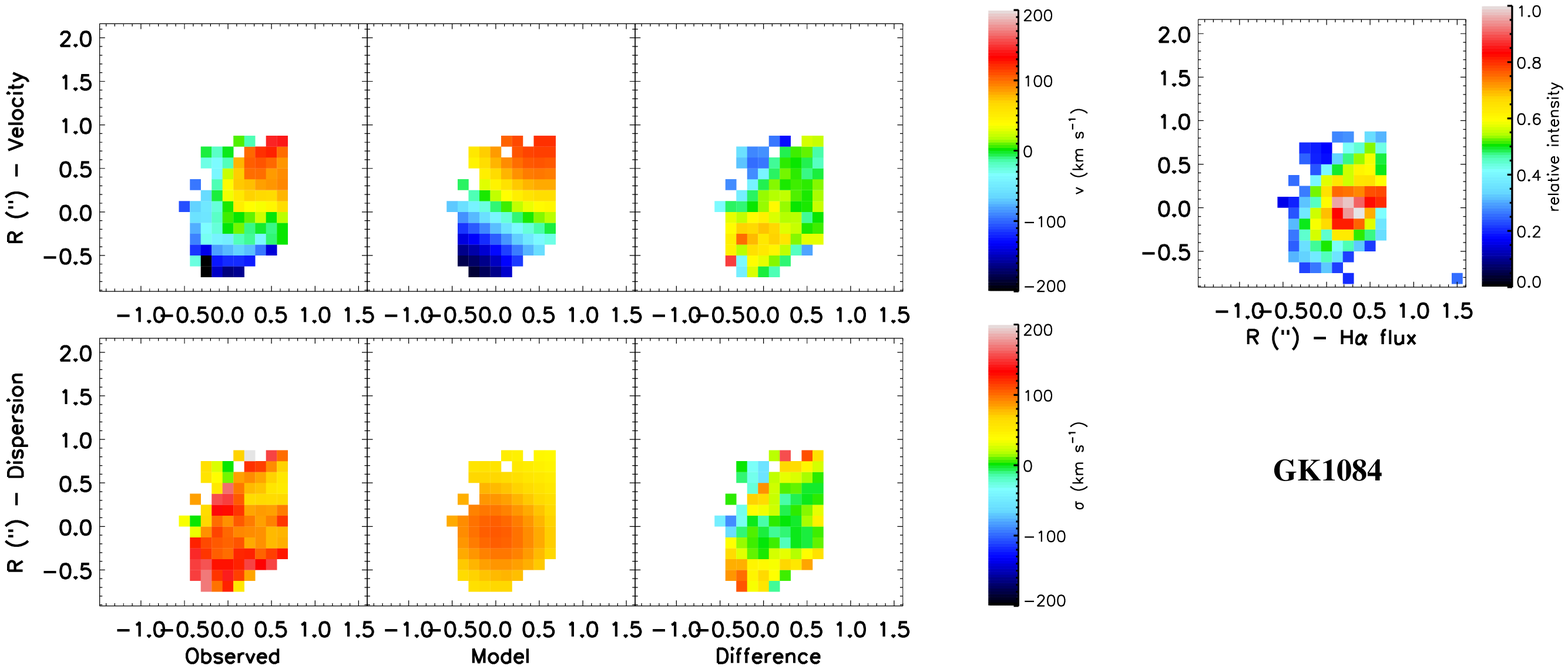}}\\
	\vspace{0.5cm}
	{\includegraphics[width=12cm,keepaspectratio]{f2_GK1084.eps}}
	
	\caption{(continued) -- Kinematic fitting of the galaxies of the sample.}
\end{figure}

	%\resizebox{\hsize}{!}{\includegraphics[clip]{panel_6397.eps}}   
	%\caption{Example of Kinematic fitting for the galaxies of the sample. For both 
	%the velocity (upper panel) and velocity dispersion (lower panel), from left 
	%to right the observed maps, the best fitting model maps and their 
	%differences. It can be seen how the global 
	%rotational pattern is tipically well reproduced, although local deviations from 
	%the ideal model are highlighted in the difference.}
	%\label{exfit}
%

%

\clearpage

\begin{table} 
	\begin{center}
	\begin{scriptsize}
	\begin{tabular}{l c c c c c c c c c c}
		\hline
		Galaxy & $z$ & $m_K$ & Select. & Band  & Scale & Exp. Time & $M_*$ & Age & H$\alpha$ & SFR \\
		&   &  &  &  & (arcsec) & (s) & ($10^{10} M_{\odot}$) & (Gyr) & ($10^{-17}$ cgs) & ($M_{\odot}/yr$)\\
		\hline
		%Q1623-BX455  &	2.408 & 21.56 & BX          & K & 0.125 & 12600 & 1.03 &  1.01  & 9.0  & 73  \\
		Q1623-BX447  &	2.148 & 20.55 & BX          & K & 0.125 & 14400 & 2.12 &  0.51  & 8.3  & 55  \\
		Q1623-BX663  &	2.434 & 19.92 & BX          & K & 0.125 & 28800 & 6.58 &  0.81  & 12.0   & 100 \\
		SSA22-MD41   &	2.172 & -     &	BX          & K & 0.125 & 28800 & 0.62 &  0.05  & 11.5 & 73  \\
		Q2343-BX389  &	2.174 & 20.18 & BX          & K & 0.125 & 14400 & 6.82 &  2.75  & 23   & 146 \\
		Q2343-BX610  &	2.211 & 19.21 & BX          & K & 0.125 & 21600 & 10.7 &  1.01  & 32   & 211 \\
		Q2346-BX416  &	2.241 & 20.30 &	BX          & K & 0.125 & 7200  & 2.24 &  0.40  & 11   & 75  \\
		Q2346-BX482  &	2.258 & -     &	BX          & K & 0.05  & 31800 & 2.90 &  0.40  & 21   & 146 \\
		%K20-ID5     &	2.227 &	18.72 &	K$<$20      & H & 0.125 & 12000 & $<7.18$ & 0.11& 25   & 167 \\
		K20-ID8      &	2.224 &	19.74 &	K$<$20      & K & 0.125 & 29400 & 3.25 &  0.45  & 11   & 73  \\
		K20-ID9	     &	2.036 &	19.94 &	K$<$20      & K & 0.125 & 34200 & 1.16 &  0.13  & 9.0  & 48  \\
		D3a-15504    &	2.383 &	19.20 &	BzK         & K & 0.05  & 20400 & 10.9 & 0.45 & 27  & 214 \\
		D3a-6004     &	2.387 &	37.63 &	BzK         & K & 0.125 & 36000 & $<31.6$ & 0.64 & 20  & 159 \\
		D3a-7144     &	1.650 &	18.70 &	BzK         & H & 0.125 & 7200  & $<11.9$ &  0.40  & 11   & 34  \\
		D3a-4751     &	2.266 &	18.94 &	BzK         & K & 0.125 & 3600  & 1.83 &  0.20  & 5.8  & 41  \\
		D3a-6397     &	1.514 &	18.11 &	BzK         & H & 0.125 & 9600  & 11.9 &  0.20  & 21   & 57  \\
		ZC782941     &	2.182 &	18.94 &	BzK         & K & 0.05  & 12600 & 2.99 &  0.14  & 20   & 123 \\
		ZC1101592    &	1.404 &	19.43 &	BzK         & H & 0.125 & 3600  & 3.81 &  0.29  & 7.4  & 16  \\
		GK2471	     &	2.430 &	20.93 &	4.5 $\mu m$ & K & 0.125 & 23400 & 1.57 &  0.13  & 4.3  & 35  \\
		GK1084	     &	1.552 &	19.31 &	4.5 $\mu m$ & H & 0.125 & 13200 & 3.61 &  0.08  & 5.0  & 13  \\
		\hline
	\end{tabular}
	\end{scriptsize}
	\caption{The galaxy sample selected for the dynamical modeling and 
	their properties. $z$ is the redshift derived from the H$\alpha$ emission 
	line, $m_K$ the apparent K band magnitude, \textit{Select.} shows the 
	criteria used to select the galaxy (see Sect.~\ref{samplesel}), 
	\textit{Scale} and \textit{Exp. time} the SINFONI pixel scale and 
	integration time used for the observations, $M_*$ and \textit{Age} 
	are the stellar mass and age of the stellar population 
	derived from the SED fitting of the sources (see Sect.~\ref{stmass}), 
	H$\alpha$ is the total line flux measured in the SINFONI data and 
	\textit{SFR} is the star formation rate derived according to Kennicutt et al. (\citealp{kennicutt}) 
	from the H$\alpha$ flux, corrected for a fixed extinction of $A_V=0.8$ 
	using a Calzetti et al. (\citealp{calzetti}) law  and for a Chabrier IMF 
	(see F\"orster Schreiber et al. \citealp{survey} for details).\label{sampletab}}
	\end{center}
\end{table}

\begin{table} 
	\begin{center} 
		\begin{tabular}{l c c c c c c}% c} 
		\hline
		Galaxy       & $i$         & $M_{dyn}(10\ kpc)$        & $M_{dyn}(R_d)$ & $\sigma_{0}$     & $R_d$ & $V_{max}$ \\%& $\chi^2$ \\
		             &             & ($10^{10} M_{\odot})$ & ($10^{10} M_{\odot})$ & (km/s)    & (kpc) & (km/s)    \\%&          \\
		\hline
		%Q1623-BX447  &	46 (+6-4)    & 11.9 (+0.6-1)    & 6.53  & 63 (+16-13)   & 6.22 & 229   \\%& 14.0 \\
		Q1623-BX447  &	46 (+6-4)    & 11.9 (+0.6-1)    & 6.5   & -  		& 6.22 & 229   \\%& 14.0 \\
		Q1623-BX663  &	28 (+6-6)    & 13.0 (+0.8-1)    & 7.5   & 40 (+24-26)   & 6.49 & 243   \\%& 6.53 \\
		SSA22-MD41   &	49 (+16-5)   & 6.8  (+1.9-0.4)  & 3.5   & 78 (+7-12)    & 5.87 & 174   \\%& 5.99 \\
		%Q2343-BX389 &	76 (+8-24)   & 13.9 (+0.2-0.2)  & 7.7   & 61 (+18-17)   & 5.96 & 259   \\%& 18.4 \\
		Q2343-BX389  &	76 (+8-24)   & 13.9 (+0.6-0.9)  & 7.7   & -		& 6.21 & 259   \\%& 18.4 \\
		Q2343-BX610  &	33 (+6-3)    & 23.4 (+9.7-0.5)  & 11    & 68 (+14-18)   & 5.38 & 324   \\%& 15.6 \\
		%Q2346-BX416 &	21 (+2-3)    & 7.7 (+0.1-0.4)   & 7.7   & 37 (+10-28)   & 1.90 & 236   \\%& 10.9 \\
		Q2346-BX416  &	21 (+7-3)    & 7.7 (+0.5-0.4)   & 2.0   & -		& 1.90 & 236   \\%& 10.9 \\
		Q2346-BX482  &	51 (+11-8)   & 12.3 (+3.5-0.7)  & 7.1   & 52 (+13-21)    & 6.35 & 237   \\%& 14.2 \\
		%K20-ID8     &  52 (+3-2)    & 9.85 (+0.2-0.1)  & 9.85  & 43 (+21-31)   & 5.36 & 209   \\%& 19.9 \\
		K20-ID8      &  52 (+6-5)    & 9.85 (+0.4-0.3)  & 4.7   & -		& 5.36 & 209   \\%& 19.9 \\
		K20-ID9	     &	52 (+11-2)   & 7    (+2-2)      & 3.5   & 39 (+28-25)   & 7.51 & 173   \\%& 16.7 \\
		D3a-15504    &	37 (+7-6)    & 16.0 (+0.4-0.3)  & 9.0   & 44 (+24-22)   & 5.31 & 284   \\%& 24.0 \\
		D3a-6004     &	30 (+35-10)  & 16   (+5-1)      & 9.8   & 58 (+12-24)   & 6.60 & 273   \\%& 0.37 \\
		%D3a-7144    &	18 (+2-3)    & 16.1 (+0.6-0.7)  & 16.1  & 30 (+13-19)   & 5.08 & 262   \\%& 18.1 \\
		D3a-7144     &	18 (+14-3)   & 16.1 (+0.6-0.7)  & 7.0   & -		& 5.08 & 262   \\%& 18.1 \\
		D3a-4751     &	16 (+13-4)   & 4.7  (+0.6-0.3)  & 1.8   & 32 (+27-21)   & 4.11 & 147   \\%& 19.9 \\
		D3a-6397     &	38 (+7-9)    & 12   (+4-3)      & 8.4   & 41 (+16-26)   & 7.62 & 235   \\%& 4.12 \\
		ZC782941     &	39 (+4-7)    & 14   (+4-3)      & 4.8   & 68 (+12-24)   & 3.60 & 257   \\%& 1.20 \\
		ZC1101592    &	84 (+4-5)    & 15   (+0.9-1)    & 2.8   & 35 (+15-18)    & 5.70 & 257   \\%& 0.52 \\
		%GK2471	     &	70 (+18-36)  & 9    (+5-4)      & 9     & 74 (+37-45)   & 3.82 & 208   \\%& 0.22 \\
		GK2471	     &	70 (+18-36)  & 9    (+5-4)      & 3.3   & -		& 3.82 & 208   \\%& 0.22 \\
		GK1084	     &	49 (+9-5)    & 12   (+2-2)      & 4.8   & 52 (+21-26)   & 4.66 & 230   \\%& 19.4 \\
		\hline
	\end{tabular}
	\caption{The results of the dynamical fitting of the galaxies using the genetic $\chi^2$ minimization. 
	$i$ is the derived inclination of the galaxy with respect to the plane of the sky; $M_{dyn}(10\ kpc)$ is 
	the dynamical mass at 10 kpc ($\sim1.2\arcsec$), while $M_{dyn}(R_d)$ at the scale length $R_d$; 
	$\sigma_{0}$ is the isotropic velocity dispersion;
	added throughout the disk (see text. Some galaxies were too compact, or with data quality not good enough, 
	to derive a robust measure of this quantity); $R_d$ the disk scale length; $V_{max}$ the maximum velocity 
	in the best fitting disk model. The relative uncertainties on $M_{dyn}(R_d)$ are the same as on 
	$M_{dyn}(10\ kpc)$, while the ones on $V_{max}$ are 1/2 those on $M_{dyn}(10\ kpc)$.
	\label{fitresults}}
	\end{center}
\end{table}

\end{document}